\documentclass[acmsmall]{acmart}
\AtBeginDocument{%
  }

\setcopyright{cc}
\setcctype{by-nc-nd}
\acmDOI{10.1145/3808164}
\acmYear{2026}
\acmJournal{PACMSE}
\acmVolume{3}
\acmNumber{FSE}
\acmArticle{FSE157}
\acmMonth{7}
\acmSubmissionID{fse26mainb-p985-p}
\received{2026-02-08}
\received[accepted]{2026-03-24}
\usepackage{xspace}
\usepackage[ruled,vlined,linesnumbered]{algorithm2e}
\usepackage[shortlabels]{enumitem}
\usepackage{listings}
\usepackage{xcolor}
\usepackage{tikz}
\usepackage{subfigure}
\usepackage{graphicx}
\usepackage[most]{tcolorbox}
\usepackage{wrapfig}
\usepackage{romannum}
\usepackage{multirow}

\definecolor{mygreen}{rgb}{0,0.5,0}
\definecolor{myblue}{rgb}{0,0,1}
\definecolor{mymauve}{rgb}{0.58,0,0.82}
\definecolor{myblack}{rgb}{0.24,0.17,0.12}
\definecolor{awesome}{rgb}{1.0, 0.13, 0.32}
\definecolor{darkred}{rgb}{0.55, 0.0, 0.0}
\newcommand\code[1]{{\tt\small #1}}
\definecolor{dkgreen}{rgb}{0,0.6,0}
\definecolor{dkblue}{rgb}{0,0.4,0.4}
\definecolor{dkviolet}{rgb}{0.3,0,0.5}
\definecolor{dkred}{rgb}{0.6,0,0}
\lstdefinelanguage{Coq}{ 
    mathescape=true,
    texcl=false, 
    morekeywords=[1]{Section, Module, End, Require, Import, Export,
        Variable, Variables, Parameter, Parameters, Axiom, Hypothesis,
        Hypotheses, Notation, Local, Tactic, Reserved, Scope, Open, Close,
        Bind, Delimit, Definition, Let, Ltac, Fixpoint, CoFixpoint, Add,
        Morphism, Relation, Implicit, Arguments, Unset, Contextual,
        Strict, Prenex, Implicits, Inductive, CoInductive, Record,
        Structure, Canonical, Coercion, Context, Class, Global, Instance,
        Program, Infix, Theorem, Lemma, Corollary, Proposition, Fact,
        Remark, Example, Proof, Goal, Save, Qed, Defined, Hint, Resolve,
        Rewrite, View, Search, Show, Print, Printing, All, Eval, Check,
        Projections, inside, outside, Def},
    morekeywords=[2]{forall, exists, exists2, fun, fix, cofix, struct,
        match, with, end, as, in, return, let, if, is, then, else, for, of,
        nosimpl, when},
    morekeywords=[3]{Type, Prop, Set, true, false, option},
    morekeywords=[4]{pose, set, move, case, elim, apply, clear, hnf,
        intro, intros, generalize, rename, pattern, after, destruct,
        induction, using, refine, inversion, injection, rewrite, congr,
        unlock, compute, ring, field, fourier, replace, fold, unfold,
        change, cutrewrite, simpl, have, suff, wlog, suffices, without,
        loss, nat_norm, assert, cut, trivial, revert, bool_congr, nat_congr,
        symmetry, transitivity, auto, split, left, right, autorewrite},
    morekeywords=[5]{by, done, exact, reflexivity, tauto, romega, omega,
        assumption, solve, contradiction, discriminate},
    morekeywords=[6]{do, last, first, try, idtac, repeat},
    morecomment=[s]{(*}{*)},
    showstringspaces=false,
    morestring=[b]",
    morestring=[d]’,
    tabsize=3,
    extendedchars=false,
    sensitive=true,
    breaklines=false,
    basicstyle=\footnotesize,
    captionpos=b,
    columns=[l]flexible,
    identifierstyle={\ttfamily\color{black}},
    keywordstyle=[1]{\ttfamily\color{dkviolet}},
    keywordstyle=[2]{\ttfamily\color{dkgreen}},
    keywordstyle=[3]{\ttfamily\color{dkblue}},
    keywordstyle=[4]{\ttfamily\color{dkblue}},
    keywordstyle=[5]{\ttfamily\color{dkred}},
    stringstyle=\ttfamily,
    commentstyle={\ttfamily\color{dkgreen}},
    literate=
    {\\forall}{{\color{dkgreen}{$\forall\;$}}}1
    {\\exists}{{$\exists\;$}}1
    {<-}{{$\leftarrow\;$}}1
    {=>}{{$\Rightarrow\;$}}1
    {==}{{\code{==}\;}}1
    {==>}{{\code{==>}\;}}1
    {->}{{$\rightarrow\;$}}1
    {<->}{{$\leftrightarrow\;$}}1
    {<==}{{$\leq\;$}}1
    {\#}{{$^\star$}}1 
    {\\o}{{$\circ\;$}}1 
    {\@}{{$\cdot$}}1 
    {\/\\}{{$\wedge\;$}}1
    {\\\/}{{$\vee\;$}}1
    {++}{{\code{++}}}1
    {~}{{$\sim$}}1
    {\@\@}{{$@$}}1
    {\\mapsto}{{$\mapsto\;$}}1
    {\\hline}{{\rule{\linewidth}{0.5pt}}}1
}[keywords,comments,strings]

\newcommand{\tightcolorbox}[2]{%
  \begingroup
  \setlength{\fboxsep}{0pt}% no padding
  \colorbox{#1}{#2}%
  \endgroup
}

\newcommand{\ourSol}{\textsc{AutoRocq}\xspace} 
\newcommand{\rev}[1]{{#1}} % 

\newcommand{\varNoPT}{${\lnot PT}$}
\newcommand{\varNoCS}{${\lnot CS}$}
\newcommand{\varNoEF}{${\lnot EF}$}
\newcommand{\varNoHM}{${\lnot HM}$}
\newcommand{\varOne}{${Err_1}$}
\newcommand{\varThree}{${Err_3}$}
\newcommand{\varFive}{${Err_5}$}

\usepackage[most]{tcolorbox}
\newtcolorbox{promptresponsebox}[1][]{
  colback=white,               % remove background color (set to white)
  colframe=black,             % frame color (your defined dark blue)
  fonttitle=\bfseries, 
  coltitle=black,              % make title font black
  colbacktitle=gray!20,        % light gray background for title
  title={Prompt-Response Example},
  boxrule=0.5pt,               % thin border
  top=1pt,                     % reduce top padding
  bottom=1pt,                  % reduce bottom padding
  left=4pt,                    % reduce left padding
  right=4pt,                   % reduce right padding
  toptitle=0.5pt,                % reduce title top margin
  bottomtitle=0.5pt,             % reduce title bottom margin
  #1
}
\theoremstyle{definition}
\newtheorem*{definition*}{Definition}
\usetikzlibrary{arrows.meta,positioning,shapes.geometric}

\newtcolorbox{resultbox}{
  colback=gray!10,     % background color
  colframe=gray!70,    % border color
  boxrule=0.8pt,       % border thickness
  arc=4pt,             % rounded corners
  left=6pt, right=6pt, top=6pt, bottom=6pt, % padding
  enhanced
}

\newcommand{\dafny}{\textsc{Dafny}\xspace}
\newcommand{\verus}{\textsc{Verus}\xspace}
\newcommand{\viper}{\textsc{Viper}\xspace}
\newcommand{\framac}{\textsc{Frama-C}\xspace}
\newcommand{\qedcartographer}{\textsc{QEDCartographer}\xspace}
\newcommand{\proverbot}{\textsc{Proverbot9001}\xspace}
\newcommand{\qedc}{\textsc{QEDC}\xspace}
\newcommand{\pbot}{\textsc{P9001}\xspace}
\newcommand{\copra}{\textsc{Copra}\xspace}
\newcommand{\palm}{\textsc{PALM}\xspace}
\newcommand{\rango}{\textsc{Rango}\xspace}
\newcommand{\coqgym}{\textsc{CoqGym}\xspace}
\newcommand{\coqstoq}{\textsc{CoqStoq}\xspace}
\newcommand{\coqhammer}{\textsc{CoqHammer}\xspace}
\begin{document}

\pagenumbering{arabic}

\title{Agentic Verification of Software Systems}

\author{Haoxin Tu}
\affiliation{%
  \institution{National University of Singapore}
  \country{Singapore}}
\email{haoxin.tu@nus.edu.sg}

\author{Huan Zhao}
\affiliation{%
  \institution{National University of Singapore}
  \country{Singapore}}
\email{zhaohuan@comp.nus.edu.sg}

\author{Yahui Song}
\affiliation{%
  \institution{National University of Singapore}
  \country{Singapore}}
\email{yahuisong123@gmail.com}

\author{Mehtab Zafar}
\affiliation{%
  \institution{National University of Singapore}
  \country{Singapore}}
\email{mehtab@nus.edu.sg}

\author{Ruijie Meng}
\affiliation{%
  \institution{National University of Singapore}
  \country{Singapore}}
\email{mengrj.cs@gmail.com}

\author{Abhik Roychoudhury}
\affiliation{%
  \institution{National University of Singapore}
  \country{Singapore}}
\email{abhik@nus.edu.sg}

\renewcommand{\shortauthors}{Haoxin Tu, Huan Zhao, Yahui Song, Mehtab Zafar, Ruijie Meng, and Abhik Roychoudhury}

\begin{abstract}
Automatically generated code is gaining traction recently, owing to the prevalence of Large Language Models (LLMs). Further, the AlphaProof initiative has demonstrated the possibility of using AI for general mathematical reasoning. Reasoning about computer programs (software) can be accomplished via general mathematical reasoning; however, it tends to be more structured and richer in contexts. 
% The structure of the program can be used to structure the proof, and this can be potentially exploited by AI agents. 
This forms an attractive proposition, since then AI agents can be used to reason about voluminous code that gets generated by AI.

In this work, we present a first LLM agent, \ourSol, for conducting automatic verification of software systems. Unlike past works, which rely on extensive training of LLMs on proof examples, our agent learns on-the-fly and improves the proof via an iterative refinement loop. The iterative improvement of the proof is achieved by the proof agent communicating with the Rocq (formerly Coq) theorem prover to get additional context and feedback. The final result of the iteration is a proof derivation checked by the Rocq theorem prover. In this way, our proof construction involves autonomous collaboration between the proof agent and the theorem prover. This autonomy facilitates the search for proofs and decision-making in deciding on the structure of the proof tree. 
Experimental evaluation on SV-COMP benchmarks and on Linux kernel modules shows promising efficacy in achieving automated program verification. As automation in code generation becomes more widespread, we posit that our proof agent can be potentially integrated with AI coding agents to achieve a generate and validate loop, thus moving closer to the vision of {\em trusted automatic programming}. 

\end{abstract}




\maketitle

\section{Introduction}  \label{sec:introduction}

The widespread adoption of AI-generated code has transformed the landscape of software development. Today, more than 25\% of the new code at Google is generated by AI \cite{google-news}. 
Yet the increasing amounts of code often carry subtle semantic errors or vulnerabilities \cite{pearce2025asleep} that may elude human review and testing, necessitating automated reasoning on program behaviors to engender \emph{trust}. 
One promising approach to attain strong, machine-checkable evidence about program properties is formal verification, which has been pivotal in certifying high-assurance systems for decades, such as microkernels~\cite{DBLP:conf/sosp/KleinEHACDEEKNSTW09}, compilers~\cite{leroy2016compcert}, and databases~\cite{verified_db}. 

Formal verification techniques commonly (1) reduce programs and specifications to logical formulas as proof obligations, and (2) discharge those obligations to automatic or interactive theorem provers (ITPs).
Common ITPs, such as Isabelle/HOL and Rocq/Coq, automatically reduce the proof goals by applying user-supplied proof steps called \emph{tactics}.
Unfortunately, formal verification sees limited adoption in production systems, as it is an extremely time- and skill-intensive undertaking. 
Specifically, (1) capturing program behaviors formally requires significant efforts in manually annotating specifications and crafting loop invariants, and (2) generating a mathematically rigorous proof to validate verification conditions demands a high degree of expertise.
For instance, the verification of the seL4 microkernel \cite{DBLP:conf/sosp/KleinEHACDEEKNSTW09} required 22 person-years of effort, while the proofs for the CompCert compiler took 6 person-years and 100,000 lines of Rocq code---eight times longer than the implementation itself \cite{leroy2016compcert}.
Most of the specifications and proofs are meticulously crafted by the developers. Recently, advances in Large Language Model (LLM) agents have been transformative across various disciplines. In particular, LLMs have demonstrated remarkable capabilities in both code understanding \cite{ahmad2021unified} and general mathematical reasoning \cite{alpha-proof}---two foundational capabilities required by program verification.  In this work, we thus ask the following question: \emph{Can LLM agents be leveraged for automatic and end-to-end verification of real-world programs?}

%\hlt{limitations of existing approaches to lemma proving.}
%\begin{itemize}[leftmargin=1em,nosep]
%\item{\it Context-agnostic tactic generation.} xxx
%\item {\it Limited understanding of proving progress.} xxx
%\item {\it Lack of sufficient feedback loop to enhance proving.} xxx
%\end{itemize}

\smallskip
{\it State-of-the-art and what is needed.} 
Since there have been approaches to proof automation that leverage LLMs \cite{lu2025palm, rango-icse25, thakur2024context} and other machine-learning-based techniques \cite{yang2019coqgym, sanchez2024qedcartographer, sanchez2020generating} — let us understand the need for a new approach. 
First of all, it is reasonable to use machine learning models to learn and predict individual proof steps or tactics in a proof. This has been accomplished in several works, including \proverbot~\cite{sanchez2020generating}. 
It is also possible for LLMs to generate whole proofs, albeit potentially incorrect ones, and then have a repair step to correct them, if possible---an approach studied in \palm~\cite{lu2025palm}. 
\rango~\cite{rango-icse25} goes one step further. Given a sequence of proof steps, it uses an ITP for yes/no validation of the proof steps, thereby allowing the trial of different tactics.

\smallskip
{\it An agentic approach.} 
These approaches all use LLMs, but are not agentic. We aim to develop an approach where an LLM agent can understand the proof structure and autonomously seek help from a theorem prover while constructing the proof. Thus, apart from using a theorem prover for validation of proof steps, the agent can actively collaborate with a theorem prover to prove a program property. As such, we develop a proof automation agent \ourSol which acts as an \emph{interpreter} of a proof derivation, and collaborates with a theorem prover to extend/improve it. 
Specifically, \ourSol \emph{autonomously} interacts with the theorem prover ahead of tactic prediction to retrieve relevant lemmas in the context, and generates tactics smartly guided by tree-shaped proof representations.
It also incorporates feedback from the interactive theorem prover and proof histories to refine tactic generation.

\smallskip
{\it Evaluation.}
We thoroughly evaluate our approach on existing mathematical lemmas~\cite{yang2019coqgym, rango-icse25}, as well as verification conditions systematically extracted from SV-COMP programs~\cite{svcomp}.
These program-derived lemmas capture intricate code logic and properties, and are more representative of program verification tasks.
We compare \ourSol against five baselines, namely \rango~\cite{rango-icse25}, \palm~\cite{lu2025palm}, \copra~\cite{thakur2024context}, \qedcartographer~\cite{sanchez2024qedcartographer}, and \proverbot~\cite{sanchez2020generating}, and results show that \ourSol significantly outperforms these state-of-the-art approaches. Specifically, \ourSol is capable of proving 48.0\% mathematical lemmas and 30.9\% verification lemmas, exceeding baseline approaches by 15.2\% to 172.8\% on mathematical lemmas, and by 10.6\% to 204.6\% on verification lemmas.
Among the successes, 168 lemmas (140 mathematical lemmas and 28 verification lemmas) are uniquely proved by \ourSol, due to its agentic design and access to proof contexts.
We also conduct a case study to verify the code in Linux kernel modules (i.e., memory management utilities), where \ourSol automatically verifies 12 lemmas for the function correctness property, compared to only 2--10 lemmas verified by baseline approaches.

\smallskip
{\it Contributions.}
In summary, the contributions of this paper are as follows.
\begin{itemize}[leftmargin=*]
    \item We build and make available\footnote{The tool is released at \url{https://github.com/NUS-Program-Verification/AutoRocq}.} 
    the first proof automation agent, \ourSol, which is highly effective in automatic proof generation. It acts as an interpreter of proof representations and actively collaborates with the Rocq prover to construct proofs.
    \item We showcase the feasibility of automatic and end-to-end verification with LLM agents. \ourSol is evaluated on the widely used SV-COMP programs in software verification, as well as Linux kernel modules. 
    \item We conduct an empirical study to demonstrate that context- and structure-awareness can help LLM agents reason about software \emph{formally}.
\end{itemize}

\section{Background}  \label{sec:background}

\subsection{Formal Program Verification}

%Formal program verification uses mathematically rigorous methods to prove that a program satisfies its specifications. Unlike traditional testing, which can only expose errors for specific inputs or executions, formal verification provides logical guarantees that a program behaves correctly for all possible executions within a given semantic model. Within this landscape, a prominent paradigm is \emph{deductive verification}. In this approach, the program code is annotated with formal specifications, such as preconditions, postconditions, and loop invariants. These annotations form \emph{formal contracts} that describe the program's intended behavior at the module and function boundaries. The annotated code is then compiled into a finite set of logical \emph{proof obligations} under an explicit semantic model. The discharge of all proof obligations implies that the program satisfies its specifications. 

Formal program verification uses mathematically rigorous methods to prove that a program satisfies its specifications. Unlike traditional testing, which only checks specific executions, formal verification provides logical guarantees for all executions within a given semantic model. A prominent paradigm is \emph{deductive verification}, where a program is annotated with formal specifications such as pre/postconditions and loop invariants. These annotations define \emph{formal contracts} for the intended behavior of the program, and the annotated code is translated into logical \emph{proof obligations}. If all proof obligations are discharged, the program is verified to satisfy its specifications.

To facilitate deductive verification, a variety of tools have been developed, including \dafny~\cite{DBLP:conf/lpar/Leino10, DBLP:journals/pacmse/MisuLM024}, 
\verus~\cite{DBLP:journals/corr/abs-2303-05491}, 
\viper~\cite{DBLP:conf/vmcai/0001SS16}, and 
\framac~\cite{DBLP:journals/fac/KirchnerKPSY15}. These tools support annotating code, generating proof obligations, and attempting to discharge them. Automated solvers~\cite{de2008z3, barbosa2022cvc5} can be used to handle simple proof obligations (e.g., quantifier-free linear arithmetic \cite{ge2009complete}). 
More complex proof obligations---such as those involving non-linear reasoning, quantifier handling, or sophisticated inductive proofs---require human guidance. In such cases, \emph{interactive theorem provers} (ITPs) are used, where auxiliary \emph{lemmas} are introduced to capture intermediate properties, helping to structure and simplify the proofs of these obligations.

%Rocq (formerly Coq) proof assistant \cite{coq1996coq} is a widely-used interactive theorem prover for semi-automated validation of residual proof obligations.
%In Rocq, proofs are constructed in a top-down manner from a \emph{proof goal}---the statement of a theorem or lemma. Users iteratively apply \emph{proof tactics}, which are simple or parameterized commands that decompose the goal into zero or more subgoals and guide the construction of a proof term. This proof term is type-checked by Rocq's trusted kernel. A proof succeeds when the subgoal list becomes empty, and the output proof term is the formal certificate of validity. This certificate can be reproduced by replaying the exact sequence of tactics, referred to as the \emph{proof script}.
%Conceptually, successful proof induces a \emph{proof tree} rooted at the original proof goal. Each node represents a subgoal produced during the derivation, and each edge corresponds to a tactic application. 
%The interactive workflow of Rocq supports incremental, exploratory proof development with frequent inspection and backtracking.

Rocq (formerly Coq) proof assistant \cite{coq1996coq} is a widely used interactive theorem prover for semi-automated validation of proof obligations. In Rocq, proofs are constructed top-down from a \emph{proof goal}, i.e., the statement of a theorem or lemma. Users iteratively apply \emph{proof tactics}, which transform the current goal into zero or more subgoals and guide the construction of a proof term. This proof term is then type-checked by Rocq's trusted kernel. A proof succeeds when no subgoals remain, and the resulting proof term serves as the formal certificate of validity. The exact sequence of tactics, called the \emph{proof script}, can be replayed to reproduce the proof. Conceptually, a successful proof induces a \emph{proof tree} rooted at the original goal, where nodes are subgoals and edges are tactic applications. This interactive workflow supports incremental and exploratory proof development with frequent inspection and backtracking.

\subsection{Machine Learning for Proof Automation}

%Proof automation aims to automatically synthesize proof scripts that discharge the given proof obligations, which closes the last mile of program verification.
%Fueled by recent advances in AI, a growing body of work tackles this task with machine-learning methods.
%Early works \cite{yang2019coqgym, sanchez2020generating, first2020tactok, sanchez2023passport} often cast this problem as a sequence generation task. 
%They design smart representations of syntactic constraints~\cite{yang2019coqgym}, proof states~\cite{first2020tactok}, and fine-grained proof contexts~\cite{sanchez2023passport} to enable neural networks to generate valid tactic sequences. However, applying a tactic successfully does not always advance the proof, as it may yield equivalent or even harder subgoals.
%This mismatch limits standalone prediction networks, which fixate on next-step tactic generation without a reliable global progress signal. 
%To address this, an orthogonal line of work \cite{blaauwbroek2020tactician, sanchez2024qedcartographer} augments these networks with search strategies to guide tactic generation.  
%For instance, \qedcartographer~\cite{sanchez2024qedcartographer} extends \proverbot\cite{sanchez2020generating} with a tactic selection model, trained through reinforcement learning. 

Proof automation aims to automatically synthesize proof scripts that discharge proof obligations, addressing the last mile of program verification. Recent advances in AI have led to a growing body of learning-based approaches. Early works \cite{yang2019coqgym, sanchez2020generating, first2020tactok, sanchez2023passport} often formulate this task as sequence generation, using representations of syntactic constraints~\cite{yang2019coqgym}, proof states~\cite{first2020tactok}, and fine-grained proof contexts~\cite{sanchez2023passport} to generate tactic sequences. However, successfully applying a tactic does not always mean the proof has progressed, since it may produce equivalent or even harder subgoals. This limits standalone prediction models that focus only on next-step tactic generation without a reliable notion of global progress. To address this, another line of work \cite{blaauwbroek2020tactician, sanchez2024qedcartographer} combines tactic prediction with search strategies. For example, \qedcartographer~\cite{sanchez2024qedcartographer} extends \proverbot\cite{sanchez2020generating} with a reinforcement-learning-based tactic selection model.

%On the other hand, Large Language Models (LLMs), with their remarkable capabilities in mathematical reasoning and high-level understanding, have become a promising alternative. A recent study~\cite{lu2025palm} shows that LLMs can grasp the high-level structure of proofs, though they often stumble on rudimentary errors such as invalid references. Building on this observation, \palm first prompts the LLM to generate a complete proof script, and then applies deterministic repair mechanisms supplemented by the external tool \coqhammer~\cite{coqhammer} to patch errors. 
%\copra~\cite{thakur2024context} is an in-context-learning approach for Lean/Coq that repeatedly proposes tactics, executes them in the ITP, and gathers error feedback to iteratively construct complete proofs under a query budget.
%More recently, \rango~\cite{rango-icse25} fine-tuned an LLM as a knowledge base for tactic generation. At every step of the proof, \rango retrieves the most relevant proofs and lemmas for the current proof state based on the predefined similarity function, and then prompts its knowledge base to generate the next tactic. 

On the other hand, Large Language Models (LLMs) have emerged as a promising alternative due to their strong mathematical reasoning and high-level understanding. A recent study~\cite{lu2025palm} shows that while LLMs can capture the high-level structure of proofs, they often make basic mistakes such as invalid references. Building on this observation, \palm first prompts the LLM to generate a complete proof script, and then applies deterministic repair mechanisms together with \coqhammer~\cite{coqhammer} to fix errors. \copra~\cite{thakur2024context} adopts an in-context learning approach for Lean/Coq, iteratively proposing tactics, executing them in the ITP, and using error feedback to construct complete proofs under a query budget. More recently, \rango~\cite{rango-icse25} fine-tunes an LLM as a knowledge base for tactic generation. At each proof step, it retrieves relevant proofs and lemmas for the current proof state and then prompts the model to generate the next tactic.

%These existing works have made significant advances in automated proof generation. However, most approaches rely on models that are trained or fine-tuned on large corpora of existing proofs to guide tactic generation. A widely-used large-scale training dataset is \coqgym~\cite{yang2019coqgym}, which contains 71K human-written proofs that cover a broad spectrum of application domains such as mathematics, computer hardware, and programming languages. Models trained or fine-tuned on this dataset are effective in automatically generating proof scripts for mathematical theorems and libraries. Unfortunately, their effectiveness is reduced when applied to proofs about computer programs, which are more complex in both structure and semantics (see a more detailed comparison between existing lemmas from \coqgym and new lemmas from programs in Section \ref{sec:background:complexity}). Moreover, existing works adopt deterministic mechanisms to patch proof errors, retrieve relevant contexts, and generate subsequent tactics, preventing them from making adaptive, demand-driven decisions. For example, context retrieval is, to the best of our knowledge, based solely on similarity functions, which may not provide sufficient information for generating the next tactic. Such a {\it non-agentic} context search strategy further limits the ability of these approaches. 

These works have significantly advanced automated proof generation. However, most rely on models trained or fine-tuned on large proof corpora to guide tactic generation. A widely used dataset is \coqgym~\cite{yang2019coqgym}, which contains 71K human-written proofs spanning domains such as mathematics, hardware, and programming languages. While models trained on this dataset are effective for mathematical theorems and libraries, they are less effective for proofs about computer programs, which are often more complex in both structure and semantics (see Section \ref{sec:background:complexity}). 
%Moreover, existing approaches typically use deterministic mechanisms for error repair, context retrieval, and subsequent tactic generation, limiting their ability to make adaptive, demand-driven decisions. For example, context retrieval is, to the best of our knowledge, based mainly on similarity functions, which may not always provide sufficient information for generating the next tactic. This {\it non-agentic} strategy further constrains their effectiveness.

\section{Motivating Example}  \label{sec:motivation}

\begin{figure}[t]
	\centering
	\includegraphics[width=0.96\linewidth]{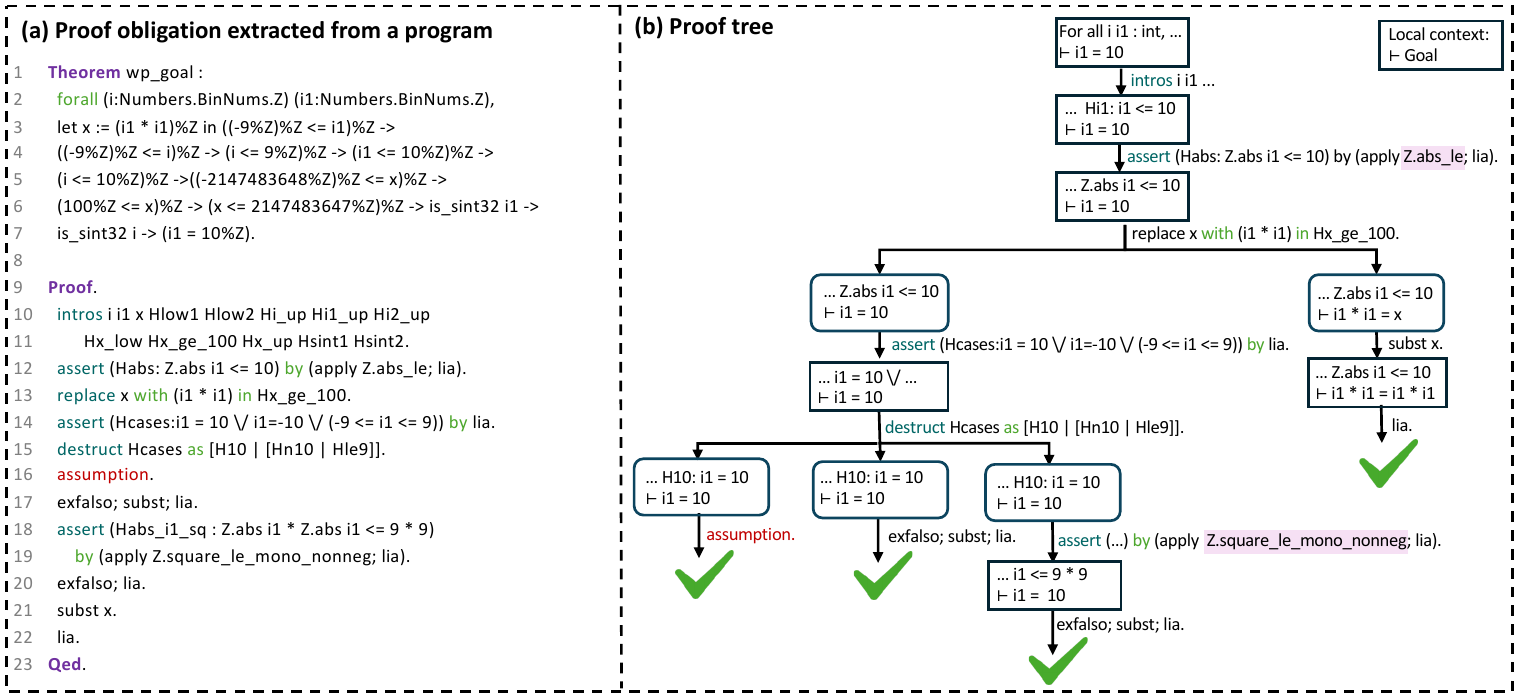}
	\vspace{-0.5em}
	\caption{(a) Proof obligation {\tt wp\_goal} extracted from {\tt benchmark52\_polynomial} in SV-COMP~\cite{svcomp} and its proof generated by \ourSol, with (b) the proof tree constructed during proving, where lemmas \tightcolorbox{pink!40!violet!20}{highlighted} are retrieved from the global context autonomously by \ourSol.}
	\vspace{-0.5em}
	\label{fig:motivating-example}
\end{figure}

Although many automatic theorem-proving approaches have been proposed, to our knowledge, none of them prove lemmas in an {\it agentic} fashion.
To motivate our agentic approach, \ourSol, we use a concrete example shown in \autoref{fig:motivating-example}.
\autoref{fig:motivating-example}(a) shows the proof obligation (\texttt{wp\_goal}, lines 1-7) extracted from verifying the {\it function correctness} property of a real-world program {\tt cggmp2005b} in SV-COMP \cite{svcomp}, along with the complete proof (lines 9-23) generated by \ourSol. 
The proof tree representation that guides \ourSol's proving process is visualized in  \autoref{fig:motivating-example}(b).
This example proof goal requires proving that ``{\it i1 = 10\%Z}~'' (line 7) holds for integer-typed ({\tt Z}) variables {\tt i1} and {\tt i} under a complex set of hypotheses involving modular arithmetic, inequalities, and type constraints (lines 2-6).
The hypotheses capture program semantics including path conditions (e.g., ``{\it i <= 10\%Z}~'') and non-overflow constraints (``{\it (-2147483648\%Z)\%Z <= x}~''), all discharged by \framac during automated verification (see \autoref{sec:lemma_extraction} for more details).
This example represents a typical lemma derived from real-world verification tasks. It captures intricate program logic and thus requires sophisticated reasoning about nested quantifiers, integer arithmetic, and logical constraints to handle.
The proof tree in \autoref{fig:motivating-example}(b) illustrates how \ourSol systematically decomposes the problem through strategic tactic application, including case analysis with {\tt destruct} that creates multiple proof branches, and maintains rich contextual information throughout the proving process.
Such complex lemmas pose significant challenges to existing approaches (e.g., \cite{rango-icse25, lu2025palm, sanchez2024qedcartographer, sanchez2020generating}), and none of them can prove this lemma during our experiments. 
We highlight these challenges to motivate \ourSol's agentic design in the following.

\smallskip
{\bf Verbose and Crowded Context.} The local context of \texttt{wp\_goal} carries rich semantic information---six hypotheses are present in the statement simultaneously. 
These hypotheses include conjunctions, disjunctions, inequalities, and type constraints. 
Conventional approaches, which mostly employ naive or similarity-based retrieval, could easily get a surfeit or an insufficiency of contextual information.
For example, before proof generation, \palm~\cite{lu2025palm} simply retrieves \emph{all} relevant premises from the global environment. In this case, it would be overwhelmed by hundreds of existing premises on integer arithmetic, most of which are irrelevant in the context. 
Conversely, \rango~\cite{rango-icse25} retrieves lemmas based on lexical similarities, an imprecise metric that often fails to identify semantically relevant lemmas. 
This is why \rango fails to prove the example: it cannot pinpoint the specific contextual assumptions on integer bounds and arithmetic properties required for the proof.

In contrast, \ourSol employs an agentic context search that  {\it autonomously}  decides, based on the current proof state, if it should make a tactic prediction, or if it should gather more context.
In the latter case, we facilitate its requests for context with fine-grained query commands (see \autoref{tab:coq-queries}).
For instance, it can leverage ``{\it Search (Z.abs \_ <= \_).}''  to fetch all lemmas and definitions that involve absolute values ({\tt Z.abs}) and match the specified wildcard pattern.
As a result, the lemma {\tt Z.abs\_le} (line 12 in \autoref{fig:motivating-example}(a)) is retrieved, which is crucial for reasoning about absolute values in the context.
This autonomy enables the agent to retrieve additional context from the theorem prover \emph{on demand}, supporting it to reason about lemmas with rich contexts.
It is worth noting that our contribution is not on how to extract certain terms or patterns in the theorem prover; rather, it is to build an agentic system that knows {\it when} to search and {\it what} patterns to search intelligently.

\smallskip
%\noindent
{\bf Understanding of Proving Progress.}
A complicated lemma such as {\tt wp\_goal} requires a delicate sequence of tactics to prove.
When generating such sequences, however, existing LLM-based methods lack structured representations of the proof process: they rely on simple goal lists rather than structured proof trees. 
This textual and linear representation of the proof state makes them overly focus on predicting a single tactic without a holistic view of the proving progress.
As a result, they often fail to adapt to the evolving state of the partially generated proof.
\ourSol explicitly maintains a well-structured proof tree representation, as shown in  \autoref{fig:motivating-example}(b).
It captures the hierarchical structure and dependencies of the proof derivation.
The proof tree is built gradually as the proving advances.
For instance, \ourSol carefully decomposes the original goal into multiple subgoals with the {\tt destruct} tactic, resulting in the branching in \autoref{fig:motivating-example}.
This structure-aware representation enables strategic reasoning about proof progress, tactical decisions, and effective tracking of proof dependencies. 
As such, our proof agent is able to \emph{interpret} the proof derivation from a higher level, thus achieving more effective proof generation.

\smallskip
%\noindent
{\bf Harnessing the Feedback.} 
Developing proofs, especially for intricate lemmas, is a trial-and-error process~\cite{shi2025qed}, and interactive theorem provers are designed to provide timely feedback to guide the proving process.
Nonetheless, existing approaches hardly leverage this opportunity to refine their tactics.
For example, \palm~\cite{lu2025palm} only performs deterministic repairs to failed tactics, whereas \rango~\cite{rango-icse25} takes only binary signals from the proof assistant and retries the prediction again.
\ourSol implements an effective feedback mechanism that incorporates the feedback from the prover to refine individual tactics. 
In addition, it also recognizes persistent errors from the history, and conducts autonomous context searches to collect additional context progressively.
This process allows our agent to learn continuously from both failures and successes. 
In this example, despite 48 failed tactic applications (59 attempts in total) during proving, \ourSol eventually recovers and synthesizes a complete proof in a short time (i.e., 156.7 seconds).

Collectively, we have an agentic proof system that can autonomously decide when and how to incorporate additional contexts and adapt strategies based on the feedback. 
A high-level interpretation of the proof derivation from the expressive proof tree supports such an agency.
In principle, this process is analogous to how an expert human prover would approach the task.
In \autoref{sec:approach}, we will explain each of these components in detail.

\section{Proof Automation with \ourSol}  \label{sec:approach}

\begin{figure}[t]
	\centering
	\includegraphics[width=0.96\linewidth]{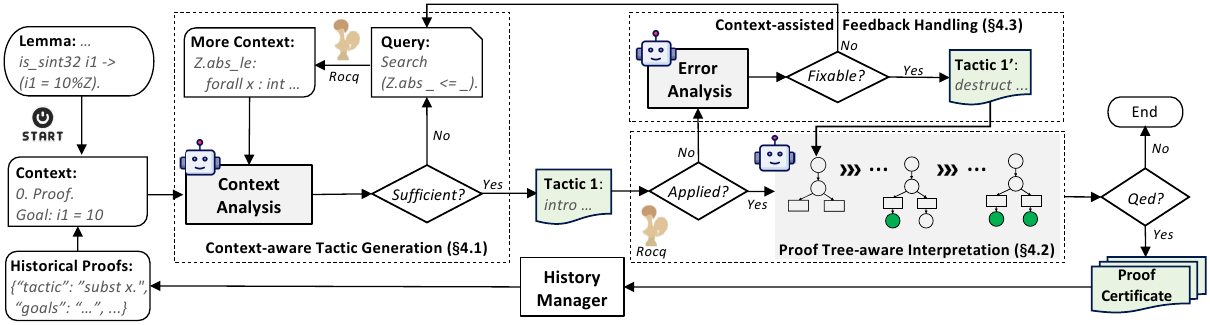}
	\vspace{-0.5em}
	\caption{Overview of \ourSol, where components involving decision-making by LLMs are \tightcolorbox{gray!20!white}{highlighted}.}
	\vspace{-1em}
	\label{fig:approach}
\end{figure}

%\smallskip
{\bf Overview.} \autoref{fig:approach} shows the overall workflow of \ourSol. 
At a high level, it is an LLM agent that takes autonomous actions to carry out a human-like proving process.
Specifically, given a lemma and the initial context (i.e., proof state), \ourSol performs an agentic context analysis to generate a tactic if the context is sufficient, or a query command if additional context is needed (\autoref{sec:approach::tactic-generation}).
Then, \ourSol generates a sequence of tactics by interpreting formal proof representations (i.e., proof tree), analyzing them, and communicating with the Rocq proof assistant autonomously (\autoref{sec:approach::proof-tree-interpretation}).
During communication, a context-assisted feedback loop is established to either refine an incorrect tactic application or provide additional context to the LLM if the errors persist, and a on-the-fly learning component is proposed to use successful proofs to help prove new lemmas (\autoref{sec:approach::feedback-handling}).
The output of a successful proof is a certificate consisting of a sequence of tactics that can be replayed in the interactive theorem prover to verify the proof.

\subsection{Context-Aware Tactic Generation}  \label{sec:approach::tactic-generation}

% This subsection describes why an agentic context-aware search mechanism is important and how \ourSol generates one tactic with the help of rich context information.

Previous works \cite{lewis2020retrieval,rango-icse25,lu2025palm, zhang2024autocoderover} have demonstrated that additional information helps LLMs generate valid tactics.
Such additional information, which we refer to as the {\it context}, includes lemmas, definitions, and existing proof steps that are relevant in constructing the proof.
Existing approaches, however, tend to augment LLMs with blindly selected context -- typically a long list of premises~\cite{lu2025palm} or existing proofs~\cite{rango-icse25} ordered by a predefined similarity metric -- without a good understanding of the current proof state or the progress of the proof.
We argue that, before suggesting a new tactic, an intelligent proof system should be able to analyze the current proof state and the progress of the proof to decide if more context is needed.
This system should carefully determine \emph{what} context information is needed and \emph{when} to add it, so that unnecessary noise is reduced to a minimum.
To support this autonomy, an agentic context-aware search mechanism is needed.
Such a context-aware approach can significantly reduce the amount of context added to the LLM, which could help the LLM focus on the current proof state and generate a more accurate tactic.

To achieve agentic context search, we couple a context analysis module with an agent that autonomously decides when and how to query the proof assistant's database.
The agent performs continuous context analysis:
if the current context is deemed sufficient, the agent directly generates a tactic to advance the proof;
if a key piece of information (e.g., the definition of a term {\tt is\_sint32} or a proved lemma in an imported library) is missing, the agent can submit a context query to retrieve it.
To retrieve the context, it constructs a precise {\it Search} command, such as ``{\it Search is\_sint32}''.   
This command is forwarded to the Rocq proof assistant, which searches its currently loaded libraries for matching identifiers, lemmas, and definitions. The results of this query are then incorporated into the LLM's context, enriching it for subsequent attempts at tactic generation.

The supported context query commands are listed in \autoref{tab:coq-queries} and the examples of query output are shown in \autoref{tab:query-examples}. By default, \ourSol supports the standard query commands available in the Rocq GUI (CoqIDE), including {\it Search}, {\it Print}, {\it Locate}, {\it About}, and {\it Check}. 
These commands enable \ourSol to retrieve critical context information on demand, significantly aiding the proof process.  
Among these, the {\it Search} command is the most frequently used (see \autoref{sec:result::rq2}). Its versatility is the key: it allows for discovery based on name patterns, type signatures, and existing lemma names~\cite{search-cmd}. This makes {\it Search} the primary tool for pinpointing relevant declarations without knowing their exact names or forms. % \emph{a priori}. 

\begin{table}[t]
\centering
\small
\caption{List of supported context query commands in \ourSol.}
\vspace{-1em}
\label{tab:coq-queries}
\begin{tabular}{lll}
\hline
\textbf{Command} & \textbf{Description} & \textbf{Output} \\
\hline
\texttt{Search <pattern>} & Search for a pattern (e.g., a term). & A list of matched declarations. \\
\texttt{Print <identifier>} & Print the definition of an identifier. & Full definition of the identifier. \\
\texttt{Locate <identifier>} & Locate where an identifier is defined. & Path or location of the identifier. \\
\texttt{About <identifier>} & Show information about an identifier. & Type and summary of the identifier. \\
\texttt{Check <term>} & Type-check a term or expression. & Type of the term or expression. \\
%\texttt{Print Assumptions <ident.>} & Show assumptions of a theorem/lemma. & List of used assumptions. \\
\hline
\end{tabular}
\vspace{-1em}
\end{table}
% preamble:
% \usepackage{multirow}

\begin{table}[t]
\centering
\small
\caption{Examples of context query commands and their outputs.}
\label{tab:query-examples}
\vspace{-1em}
\setlength{\tabcolsep}{6pt}
\renewcommand{\arraystretch}{1.15}

\begin{tabular}{p{0.3\linewidth} p{0.61\linewidth}}
\hline
\textbf{Example Command} & \textbf{Output (reduced version)} \\
\hline

\multirow{2}{*}{\texttt{Search (Z.quot \_ \_).}} &
\texttt{\small Zquot.Z\_mult\_quot\_le: forall a : int, 0 ÷ a = 0}\\
&
\texttt{\small Z.quot\_unique\_exact: \xspace forall a : int, a ÷ 0 = 0}\\
\hline

\multirow{1}{*}{\texttt{Print is\_sint32.}} &
\texttt{\small is\_sint32=fun x:int => -2147483648 <= x < 2147483648...} \\
  
\hline

\multirow{1}{*}{\texttt{Locate to\_sint64.}} 
& \texttt{\small Constant main\_assert.to\_sint64.} \\
\hline

\multirow{1}{*}{\texttt{About Z.abs.}} 
& \texttt{\small Z.abs : int -> int.} \\
\hline

\multirow{1}{*}{\texttt{Check why\_decidable\_eq.}} &  \texttt{\small why\_decidable\_eq : forall x y : bool, {x = y} + {x <> y}}\\
\hline

\end{tabular}
\vspace{-0.5em}
\end{table}

%\begin{wrapfigure}{r}{0.6\textwidth}
\vspace{-0.5em}
\begin{promptresponsebox}[title={Agentic Context Analysis}]
{\bf \textit{Prompt:}}
``Analyze the current \{proof tree\} and Top-5 \{historical tactics\}. If sufficient information is available, generate a tactic to proceed. Otherwise, output a query command to retrieve additional context.''

{\bf \textit{Response:}} \lstinline[language=Coq]{Search (Z.abs _ <= _).}
%\begin{lstlisting}[language=Coq,numbers=none]
%Search is_sint32.
%\end{lstlisting}
\end{promptresponsebox}
\vspace{-0.5em}
%\end{wrapfigure}

\subsection{Proof Tree-Aware Interpretation} \label{sec:approach::proof-tree-interpretation}

The key component empowering \ourSol's agency is an expressive proof tree representation that enables high-level interpretation of the proof derivation. 
Compared to the textual or linear representation of proof states (e.g., a list of goals), the explicit tree structure conduces understanding of the proving process.
Formally, we define a proof tree as follows \cite{lample2022hypertree, prooftree}.

\begin{definition*}[{\it Proof Tree}]
Let $\Sigma$ be the proving context\footnote{Here, $\Sigma$ captures both the global environment and the local context, as they are not differentiated in our context search.}, $G$ be the initial proof goal, and $S$ be a sequence of tactics that transforms $G$ under $\Sigma$.
A proof tree $T(\Sigma, G, S)$ is inductively defined as:
\begin{itemize} [leftmargin=1em,nosep]
    \item Each node $N_A$ corresponds to a goal $\Sigma \vdash A$, i.e., a statement $A$ to be proved under $\Sigma$.
    \item The root node is $N_G := \Sigma \vdash G$.
    \item When a tactic $\tau \in S$ is applied to a node $N_A := \Sigma \vdash A$, it either
    (1) generates zero subgoals, in which case $N_A$ is a leaf, or 
    (2) generates $n$ subgoals/nodes $\{N_{A_i} := \Sigma \vdash A_i \mid i=1,\dots,n \}$, and creates edges $N_A \rightarrow N_{A_i}$.
    \item A node is a leaf if it is directly derivable under $\Sigma$, i.e., \ no further subgoals are produced.
\end{itemize}
\end{definition*}

Note that the $S$ may or may not be a complete tactic sequence that proves $G$.
Intuitively, a proof script $S$ induces a proof tree $T$ in its enclosing context $\Sigma$, rooted at the original proof goal $G$. Each node represents a subgoal to be proved, and a tactic application is represented as an edge. Leaves are subgoals that are trivially true or could be easily proved with a single tactic.
The goal is complete when all the sub-goals in the leaves are proved.

\ourSol explicitly maintains a proof tree as it progresses in tactic generation.
After a successful tactic application, it examines the open subgoals from the Rocq proof assistant.
If the tactic successfully proves the current subgoal (i.e., a leaf node), \ourSol marks it as closed, and shifts its focus to another residual subgoal.
When the tactic creates multiple subgoals, it creates these nodes in the tree and determines one of the new subgoals to work on.
In this way, the proof tree is updated by extending the current node.
On the other hand, if the tactic fails, our agent remains at the current node on the proof tree for further attempts.
During the entire proving process, the proof states in subgoals and tactic applications are continuously updated until a complete tree is constructed.
\autoref{fig:motivating-example}(b) shows an example of a proof tree after all tactics were applied in the motivating example presented in \autoref{fig:motivating-example}(a).
From the tree, we could see that four subgoals were generated in total, and the lemmas with a violet background were retrieved from our agentic context search.
The proof is completed when all leaves are closed.

Whenever the proof tree is updated, \ourSol \emph{interprets} the tree structure comprehensively to reevaluate the current progress. 
It does so by traversing the proof tree to identify the open subgoals and how they relate to the existing proof structure.
With a holistic view of the entire derivation process, it then proceeds to generate a sequence of tactics iteratively.

We note that our key contribution is not about formulating the proof tree structure. Instead, our insight lies in the realization that high-level interpretation enables LLM agents to carry out autonomous decision-making and actions.
The benefits provided by our proof tree representation are twofold.
First, it captures the hierarchical structure of the proof, including the relationships between tactics and subgoals, which provides a richer context for interpreting the proof process.
Second, it allows \ourSol to systematically break down the proof into smaller subgoals and tackle them one by one, which is aligned with how humans approach proving \cite{subgoal-proving, shi2025qed}.

\subsection{Context-Assisted Feedback Handling} \label{sec:approach::feedback-handling}
To emulate a trial-and-error feedback loop, we augment \ourSol with feedback handling mechanisms to collaborate with the Rocq ITP. 
Based on the error that occurs, \ourSol employs two strategies during tactic generation: (1) revising the tactic based on the error message when a single error occurs, and (2) initiating a context search to retrieve additional context when persistent errors occur. 
In addition, \ourSol takes positive feedback from past successes. We delegate each mechanism to a subsection as follows.

\subsubsection{Handling a Single Error with Strategic Fixing}
When a tactic fails to apply to a subgoal, the ITP provides an error message indicating the reason for the failure.
As shown in the following example, \ourSol captures this error message and feeds it back to the LLM, along with the current proof tree.
The LLM then analyzes the error message and the context to generate a revised tactic to fix the issue.
Subsequently, the revised tactic is applied again, and the proof tree is updated accordingly.

%\begin{wrapfigure}{r}{0.6\textwidth}
%\vspace{-0.5em}
\begin{promptresponsebox}[title={Agentic Error Handling of a Single Error}]
{\bf \textit{Prompt:}}
``The previous \{tactic\} failed to apply to the current subgoal with the following \{error message\} from Rocq: \{error message\}.
Analyze the error and generate a corrected tactic.''

{\bf \textit{Response:}} \lstinline[language=Coq]{destruct Hcases as [H10 | [Hn10 | Hle9]].}
%\begin{lstlisting}[language=Coq,numbers=none]
%destruct Hcases as [H10 | [Hn10 | Hle9]].
%\end{lstlisting}
\end{promptresponsebox}
%\vspace{-0.5em}
%\end{wrapfigure}

\subsubsection{Handling Persistent Errors with Context Search}
If the same error persists after several repair attempts (which is treated as {\it not fixable} as shown in Figure \ref{fig:approach}), \ourSol recognizes that the current context may be insufficient to resolve the subgoal.
In such cases, \ourSol initiates a context search by issuing a query command to the ITP to retrieve additional context information, similar to the process described in \autoref{sec:approach::tactic-generation}. 
The retrieved context is then incorporated into the LLM's input, and the LLM generates a new tactic based on the enriched context.
This context-assisted feedback loop continues until the subgoal is successfully solved or a predefined termination condition is met (e.g., maximum number of attempts).
For example, the following shows how \ourSol handles persistent errors. 
Given the failed tactics and the proof tree, our agent returns a query command to search for what is defined in the integer quotient from the {\tt Z.quot} module.

%\begin{wrapfigure}{r}{0.6\textwidth}
%\vspace{-0.5em}
\begin{promptresponsebox}[title={Agentic Error Handling of Persistent Errors}]
{\bf \textit{Prompt:}}
``The agent has repeatedly generated \{failed tactics\} for the current subgoal multiple times. 
Analyze the current \{proof tree\} and determine what additional context is needed to proceed.
Output a query command to retrieve the necessary context information.''

{\bf \textit{Response:}} \lstinline[language=Coq]{Search (Z.quot _ _).}
%\begin{lstlisting}[language=Coq,numbers=none]
%Search (Z.quot _ _).
%\end{lstlisting}
\end{promptresponsebox}
%\vspace{-0.5em}
%\end{wrapfigure}

\subsubsection{Managing Historical Proofs for On-the-fly Learning}
\ourSol learns on the fly by maintaining an archive of successful proofs. This archive is accumulated throughout a proving campaign, and \ourSol adapts its strategy when proving new lemmas.
Specifically, when a lemma is successfully proved, \ourSol stores comprehensive tactic history records that capture the complete context and evolution of each proof step. 
Each record contains the applied tactic, the proof goals before and after tactic application, the available hypotheses and their changes, along with metadata including the theorem name, and the tactic ID within the proof sequence. 
This rich historical information serves multiple purposes: (1) it enables \ourSol to learn from successful proof patterns and reuse effective tactic sequences in similar contexts, (2) it provides detailed examples for context-aware tactic generation by showing how specific goals were transformed, and (3) it supports the feedback handling mechanism by offering concrete instances of successful problem-solving strategies. 
By maintaining this detailed provenance of proof construction, \ourSol can build a knowledge base of proven tactics that enhances its capability to tackle new lemmas with similar structural patterns or mathematical properties (the experiment results presented in Section \ref{sec:result::rq2} also support our claim).

%\smallskip
%\noindent
%{\bf Implementation.}
\subsection{Implementation}
We implemented \ourSol in approximately 8k lines of Python code with a modular architecture that separates the core proving logic, theorem prover interface, and supporting utilities. The main Rocq backend is implemented using the CoqPyt (v1.0.0) library \cite{carrott2024coqpyt} to interact with the Rocq proof assistant.
The implementation follows an iterative workflow as shown in \autoref{fig:approach}.
The modular design enables easy extension and maintenance while supporting the three key components of our approach: context-aware tactic generation, proof tree management, and feedback handling.
Historical proof data is stored in JSON format to enable learning from successful proof patterns for future lemmas.
For the LLM backend, we use GPT-4.1 as the backbone model for all LLM interactions. We set the temperature to 0 for reproducibility.

% \subsection{Implementation of \ourSol}

\section{Experimental Setup}  \label{sec:setup}

To evaluate the effectiveness of \ourSol on program verification tasks, we seek to answer the following research questions (RQs):

\begin{description}
    \item [\textbf{RQ.1}:]  Is \ourSol effective at proving mathematical lemmas from \coqgym? 
    \item [\textbf{RQ.2}:]  Is \ourSol effective at proving lemmas from program verification tasks?
    \item [\textbf{RQ.3}:] How does each component of \ourSol contribute to its effectiveness?
    \item [\textbf{RQ.4}:] How do the proofs generated by \ourSol compare with human written proofs?
\end{description}
In this section, we first present our experimental setup.
We then analyze the detailed results in Section \ref{sec:result}.
We also conduct case studies on verifying individual Linux kernel modules in \autoref{sec:linux}.

\subsection{Preparation of Benchmark Lemmas} \label{sec:benchmark_preparation}

\subsubsection{Lemmas from Existing Datasets} \label{sec:benchmark_existing}
\coqgym~\cite{yang2019coqgym} and \coqstoq~\cite{rango-icse25} are two well-known datasets for evaluating neural theorem provers in Rocq, and have been extensively studied in the community~\cite{blaauwbroek2020tactician, lu2025palm, rango-icse25}.
For fair and comprehensive evaluation, we include only the lemmas that come from the {\it intersection} of \coqstoq and \coqgym testing sets. This is because other projects from \coqgym's test set are included in \coqstoq's training set, and thus may cause inflated results.
Specifically, these include seven projects (i.e., {\tt dblib}, {\tt zfc}, {\tt hoare-tut}, {\tt huffman}, {\tt buchberger}, {\tt PolTac}, and {\tt zorns-lemma}), which represent common mathematical and human-written lemmas, such as logical tautology, coding algorithms, and theory formalizations.
In total, there are 1,717 mathematical lemmas selected to evaluate comparative approaches.

\subsubsection{Lemmas for Program Verification Tasks.} \label{sec:benchmark_selection}

While \coqgym~\cite{yang2019coqgym} and \coqstoq~\cite{rango-icse25} provide a valuable corpus of human-written theorems, they are drawn primarily from mathematical libraries and do not capture the complexity of program verification lemmas found in real-world software. Since \ourSol is designed to automate program verification, its evaluation requires a benchmark of lemmas extracted from real programs. To our knowledge, no such dataset currently exists, leaving a critical gap in the program verification field.

To address this gap, we construct a new benchmark by systematically extracting lemmas from real-world C programs. This benchmark enables the evaluation of \ourSol and facilitates future research in automatic program verification.
We source our subject programs from SV-COMP~\cite{svcomp} for two key reasons: their widespread adoption in the formal verification community ensures they represent a diverse set of C language constructs, and their inclusion of ground-truth specifications provides meaningful proof obligations.
Our selection focused on the most common property types in SV-COMP: functional correctness, defined by the unreachability of error calls, and non-overflow. The final criteria mandated that a program must (1) be verified for both properties and (2) exhibit deterministic behavior (e.g., no multi-threading). 
Applying these criteria yielded a final benchmark of 131 C programs from SV-COMP. The programs have an average of 43.06 lines of code, with the largest (the utility {\tt basename} in BusyBox \cite{busybox}) comprising 428 lines.

\subsubsection{Lemma Extraction Methodology} \label{sec:lemma_extraction}

We designed a systematic method to automatically extract lemmas from the selected SV-COMP programs. We focus on generating non-trivial lemmas that necessitate reasoning about program semantics, as opposed to simple lemmas that can be discharged with basic tactics (e.g., {\tt auto.}). Our approach utilizes \framac~\cite{DBLP:journals/fac/KirchnerKPSY15} to generate the necessary proof obligations from the program code.

Technically, we first use the RTE plug-in~\cite{rte} to annotate the source code with formal contracts in the ANSI/ISO C Specification Language (ACSL), including preconditions and postconditions. 
As \framac cannot automatically infer loop invariants, we address this by employing LLMs to generate candidate invariants based on the loop's context and structure. Each candidate is validated through property testing with the Eva plug-in~\cite{eva}; unsuccessful candidates are refined iteratively until a verifiably correct invariant is established.
Note that while property testing can falsify incorrect invariants by discovering counterexamples, it cannot formally prove the correctness of an invariant -- that it holds for all possible program executions. 
We justify this design choice as loop invariant inference is a long-standing, challenging problem~\cite{specgen-icse25,learning-invariants} that is orthogonal to our core focus on proof automation. Once the code is fully annotated and verified, we employ \framac's WP plug-in~\cite{wp} to generate proof obligations. These obligations are automatically translated into lemmas, which can be discharged in the Rocq proof assistant. 
The overall framework completes an end-to-end automated workflow from C programs to verifiable Rocq lemmas. 

As a result, we collected 641 lemmas extracted from the verification of two types of target properties: non-overflow and functional correctness.
Since both property types often require reasoning about loops, a significant portion of the generated proof obligations are related to loop invariants. These are essential intermediate lemmas needed to conclude that either a non-overflow or a functional correctness property holds for the entire program.
Consequently, the 641 lemmas comprise three categories, reflecting their role in the proof: (1) non-overflow lemmas (203, 31.7\%): direct proof obligations for non-overflow properties; (2) functional correctness lemmas (153, 23.9\%): direct proof obligations for functional properties; 
and (3) loop invariant lemmas (285, 44.5\%): Supporting lemmas required to prove the loop invariants necessary for establishing both non-overflow and functional correctness properties. 
Together, these lemmas represent a diverse set of challenging proof obligations that arise in real-world program verification tasks.

\subsubsection{Comparison of Extracted Lemmas and Existing Lemmas} \label{sec:background:complexity}

To quantify the complexity and difficulty of verification lemmas extracted in \autoref{sec:lemma_extraction}, we compare them against existing mathematical lemmas pooled in \autoref{sec:benchmark_existing}.
To do this, one commonly used proxy is the difficulty of proving a lemma, since more sophisticated lemmas generally necessitate longer proofs involving more cases and derivation steps. 
Unfortunately, this is infeasible for our purpose, since no ground-truth proofs are readily available for our extracted lemmas. As such, we directly examine these lemmas themselves.
Specifically, we propose two metrics: (a) term count: the number of terms (variables, quantifiers, operators) to gauge structural complexity, and (b) hypothesis count:  the number of assumptions to measure the richness of the semantic context.

\begin{figure}[tbp]
  \centering
  % \subfigure[Lemma Length]{
  %   \includegraphics[width=0.3\linewidth]{figures/length.pdf}
  %   \label{fig:length}
  % }
  %\hfill
  \subfigure[Term Count]{
    \includegraphics[width=0.38\linewidth]{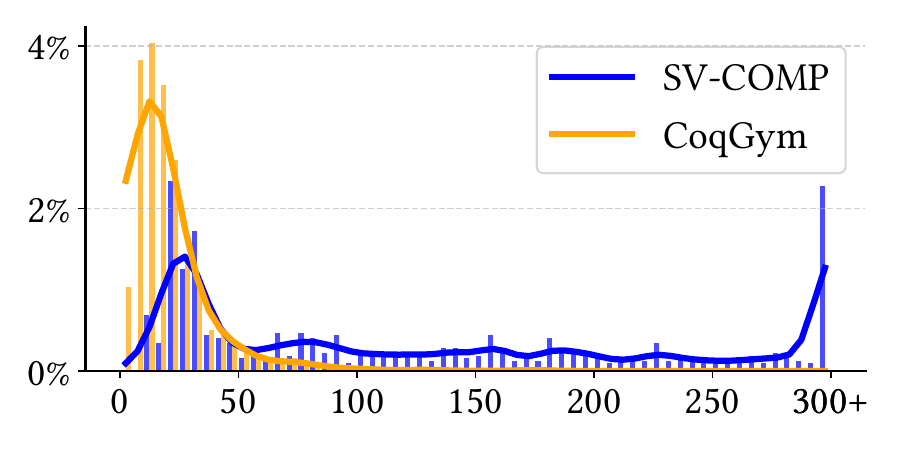}
    \label{fig:term}
  }
  %\hfill
  \subfigure[Hypothesis Count]{
    \includegraphics[width=0.38\linewidth]{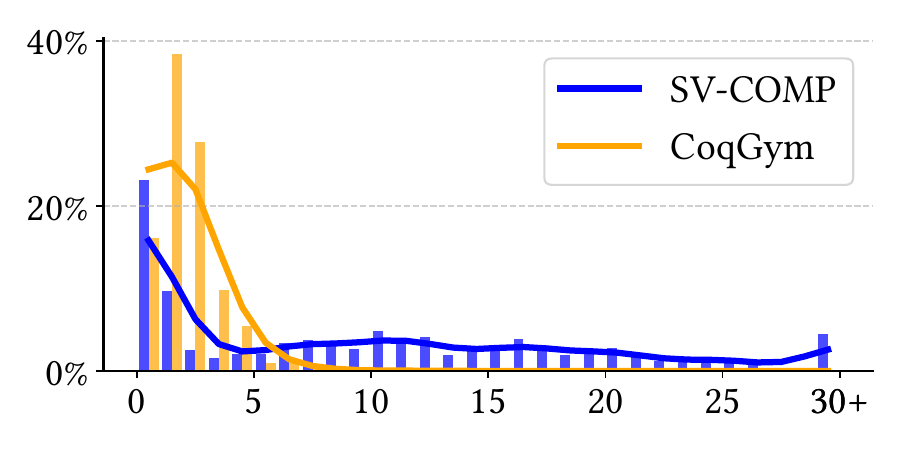}
    \label{fig:hypothesis}
  }
  % \hfill
  % \subfigure[Quntifier Count]{
  %   \includegraphics[width=0.23\linewidth]{figures/quant.pdf}
  %   \label{fig:quant}
  % }
  %\vspace{-1em}
  \caption{Histogram of different complexity metrics: lemmas from SV-COMP programs (blue) vs. \coqgym (orange). \rev{Less than 1\% of lemmas from \coqgym involve more than 100 terms or 7 hypotheses.}}
  \label{fig:complexity}
  \vspace{-0.5em}
\end{figure}

In \autoref{fig:complexity}, we report the complexity distribution of the lemmas by measuring the percentage of lemmas (y-axis) with a given complexity metric (term or hypothesis count, x-axis). 
For both metrics, the lemmas from \coqgym (orange) densely concentrate on the left end of the axis, suggesting uniformly lower complexity. 
In contrast, lemmas extracted from SV-COMP programs (blue) exhibit a more even distribution, with a lower peak and a long right tail, reflecting generally higher complexity.
Concretely, 48\% of these lemmas from SV-COMP programs consist of >100 terms, and 52\% of them are associated with >7 hypotheses; $<1\%$ mathematical lemmas satisfy either constraint.
The stark contrast across both metrics suggests that the obligations extracted from real-world programs, which often capture intricate logic in the code, tend to have much higher complexity than their benchmark counterparts.
Below shows an example lemma from {\tt hoare\_tut}, a project in \coqgym. It states that the non-equality function {\tt Zneq\_bool} correctly returns false only if {\tt x = y}. It is much simpler than lemmas from program verification tasks, as shown in \autoref{fig:motivating-example}(a).

\begin{lstlisting}[language=Coq,basicstyle=\scriptsize,numbers=none]
Lemma Zneq_bool_false: forall x y, Zneq_bool x y=false -> x=y.
\end{lstlisting}

% {\bf Benchmarks used in Evaluation.}
In summary, we use the two benchmark sets for our evaluation in \autoref{sec:result}, namely (1) 1,717 mathematical lemmas from the intersection of the test partition of \coqgym~\cite{yang2019coqgym} and \coqstoq~\cite{rango-icse25}, and (2) 641 verification-derived lemmas from SV-COMP programs~\cite{svcomp}.
In total, we have 2,358 lemmas across different domains.

\subsection{Comparative Baselines}

We compare \ourSol with state-of-the-art tools developed for proof automation in Rocq. 
We choose five baselines, namely \rango~\cite{rango-icse25}, \palm~\cite{lu2025palm}, \copra~\cite{thakur2024context}, \qedcartographer~\cite{sanchez2024qedcartographer} (or \qedc for short), and \proverbot~\cite{sanchez2020generating} (or \pbot for short). 
These tools have shown exceptional results on proof automation tasks, and they employ different machine-learning techniques: standard LLMs, fine-tuned LLMs, reinforcement learning, and recurrent neural networks (RNNs), respectively.
\ourSol, \palm, and \copra, which invoke closed-source LLM through APIs, use GPT-4.1 as the backend with temperature 0.
For the other three baselines, we use the pre-trained weights made available in their artifacts.
These include a fine-tuned version of DeepSeek-Coder used by \rango, a reinforcement-learning-based tactic selection agent employed by \qedc, and a custom tactic prediction network in \pbot.
They also have access to a single Nvidia A40 GPU with 48GB of VRAM if needed.
We use the default settings for all the tools. \rango and \copra times out after 10 minutes, and \qedc is limited to 512 generation steps. \palm and \pbot are executed until completion.
We set up all the baselines in Docker on Ubuntu 22.04.
Other than \palm, which uses older Rocq 8.12 for compatibility reasons, all the other tools are configured with Rocq 8.18. 
We delay the discussions on the use of different models and Rocq versions to Section~\ref{sec:threat}.

\section{Experimental Results} \label{sec:result}

\subsection{RQ.1: Efficacy of \ourSol on Mathematical Lemmas} \label{sec:result::rq1a}

\begin{figure}[t]
\begin{minipage}[t]{0.48\linewidth}
  \centering
    \includegraphics[width=\linewidth]{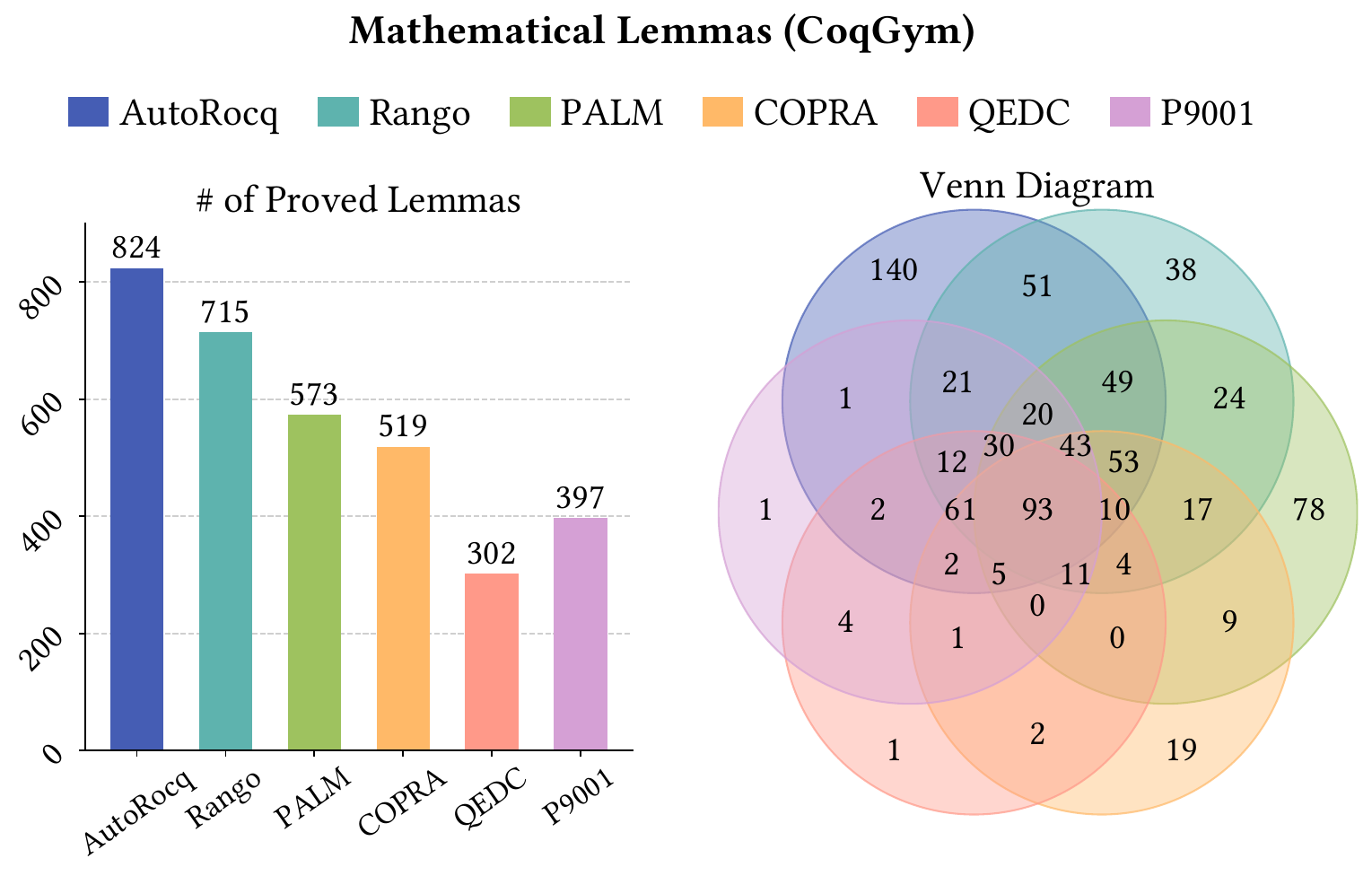}
  \vspace{-2em}
  \captionof{figure}{[RQ.1] Mathematical lemmas from \coqgym proved by each tool and their breakdown. }
  \label{fig:rq1a}
  \vspace{-1em}
\end{minipage}\hfill
\begin{minipage}[t]{0.48\linewidth}
  \centering
  \includegraphics[width=\linewidth]{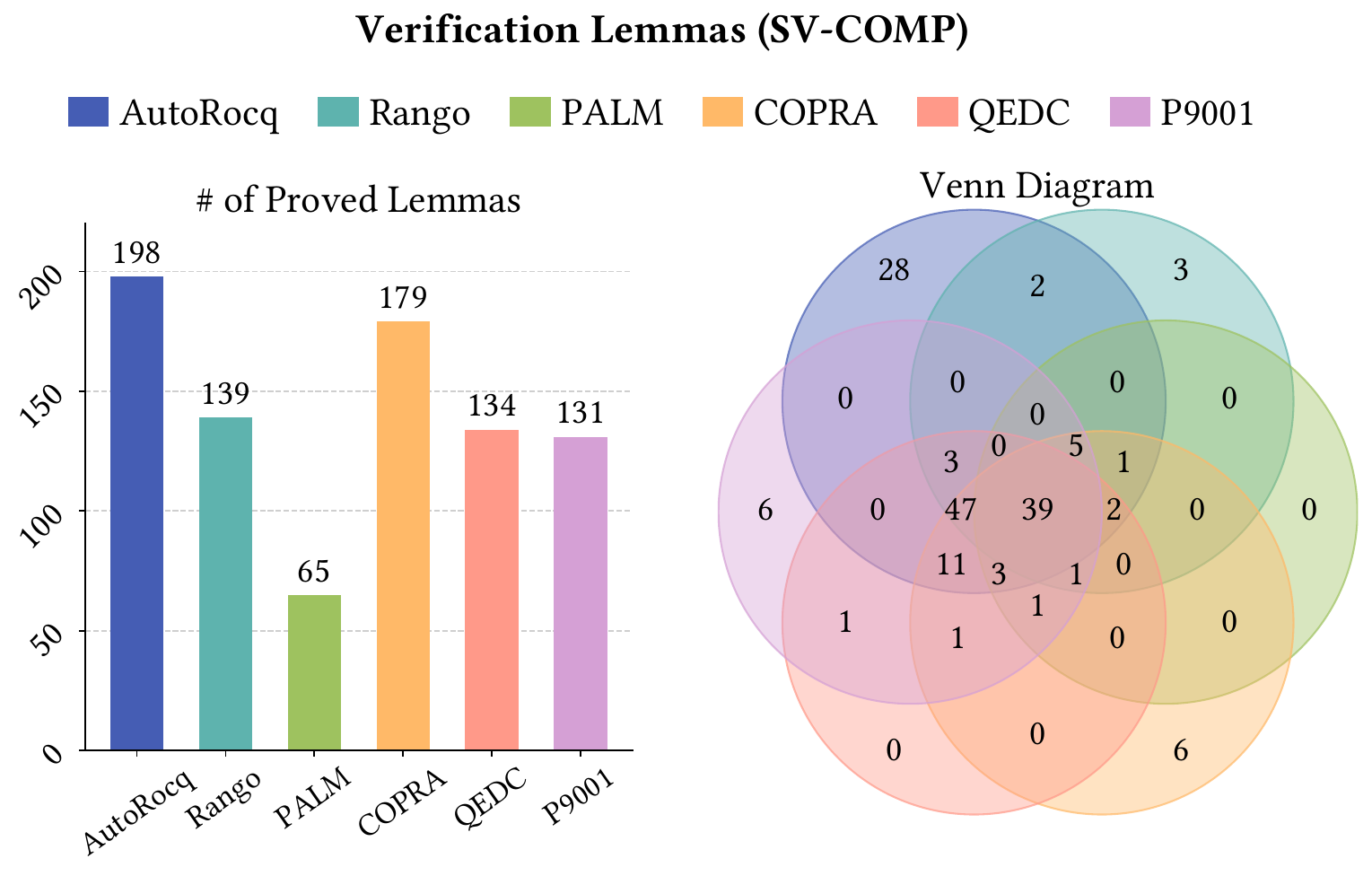}
  \vspace{-2em}
  \captionof{figure}{[RQ.2] Verification lemmas from SV-COMP programs proved by each tool and their breakdown. }
  \label{fig:rq1b}
  \vspace{-1em}
\end{minipage}
\end{figure}

For this research question, we investigate \ourSol's effectiveness in proof generation on mathematical lemmas.
We report the number of proved lemmas from each tool and a detailed breakdown in \autoref{fig:rq1a}. 
The bar plot (left) shows the total number of lemmas successfully proved by each tool, whereas the Venn diagram (right) presents the detailed breakdown of the proved lemmas.
We note that \palm's results may not reflect its full capability, as several lemmas in the benchmarks are incompatible with Rocq 8.12.

The bar plot (\autoref{fig:rq1a}, left) shows that, across all 1,717 lemmas, \ourSol successfully proves 824 (48.0\%) of them, outperforming other tools by a large margin of 15.2\% (compared with \rango) to 172.8\% (compared with \qedc). 
Among the baseline tools, \rango performs the best on this dataset, synthesizing proofs for 715 (41.6\%) of the lemmas on \coqgym. 
The two baselines using GPT-4.1 as the backend model, \palm and \copra, can both prove more than 500 mathematical lemmas. 
The Venn diagram (\autoref{fig:rq1a}, right) also demonstrates that \ourSol proves the most lemmas (140) uniquely. Our tool is followed by \palm and \rango, which account for 78 and 38 uniquely proved lemmas, respectively.
\copra is able to find proofs for 19 lemmas uniquely.
The results suggest that \ourSol is highly effective in automatic proof generation.

%\noindent

\subsection{RQ.2: Efficacy of \ourSol on Verification Lemmas} \label{sec:result::rq1b}

More importantly, we study complex proof obligations that represent real verification workflows, and present the results in \autoref{fig:rq1b}, with the bar plot (left) and its breakdown in the form of a Venn diagram (right).
Overall, \ourSol manages to prove 198 lemmas (30.9\%) from the benchmark, outperforming the best baseline, \copra, by 10.6\%.
Results produced by \rango, \qedc, and \pbot lie in close proximity to one another, with each producing proofs for around 135 lemmas.
\ourSol outperforms them by 42.4\%, 47.8\%, and 51.1\%, respectively. 
The number of lemmas proved by \ourSol is also 204.6\% higher than that of \palm. 

The Venn diagram (\autoref{fig:rq1b}, right) shows that \ourSol generates proof scripts for 28 unique lemmas (owing to its agentic workflow) that \emph{no} other tool can prove.
In contrast, \emph{no} other tool is able to generate proofs uniquely for more than 6 lemmas.
In fact, only 19 lemmas proved by \emph{any} other baselines elude \ourSol!
We note that there is a large overlap among the successes of baseline tools. 113 lemmas are proved commonly by at least three of the baselines, suggesting a strong convergence in capabilities among them.

Compared to the results in \autoref{fig:rq1a}, the verification lemmas indeed pose a greater challenge to \emph{all} neural theorem provers (including \ourSol), as evidenced by their lower success rates overall. 
The relative degradation in performance is the most significant for \palm, indicating that whole-proof generation scales poorly as the lemmas grow in complexity. 
Similarly, \rango is the best-performing baseline on mathematical lemmas, but is outperformed by \copra in the SV-COMP benchmark. This performance decline can be attributed to a distribution mismatch: its underlying model was fine-tuned primarily on \coqstoq, and consequently struggles when encountering the more complex verification lemmas that lie outside this training data.
In \emph{both} benchmarks, \ourSol proves the most lemmas \emph{and} has the most unique successes, strongly suggesting its efficacy and robustness in automated proof generation.

\begin{resultbox}
{\bf Answer to RQ.1\&2}: Overall, \ourSol is more effective in proving both mathematical and verification-derived lemmas, outperforming baselines by 15.2\% to 172.8\% on \coqgym and 10.6\%--204.6\% on SV-COMP. 
Notably, \ourSol proves 168 lemmas (140 on \coqgym and 28 on SV-COMP) that none of the other approaches can prove.
\end{resultbox}

\begin{figure}[t]
%\vspace{-2em}
\centering
\subfigure[Breakdown by term count.]{	
\begin{minipage}[t]{0.48\linewidth}
\centering
\includegraphics[width=1\linewidth]{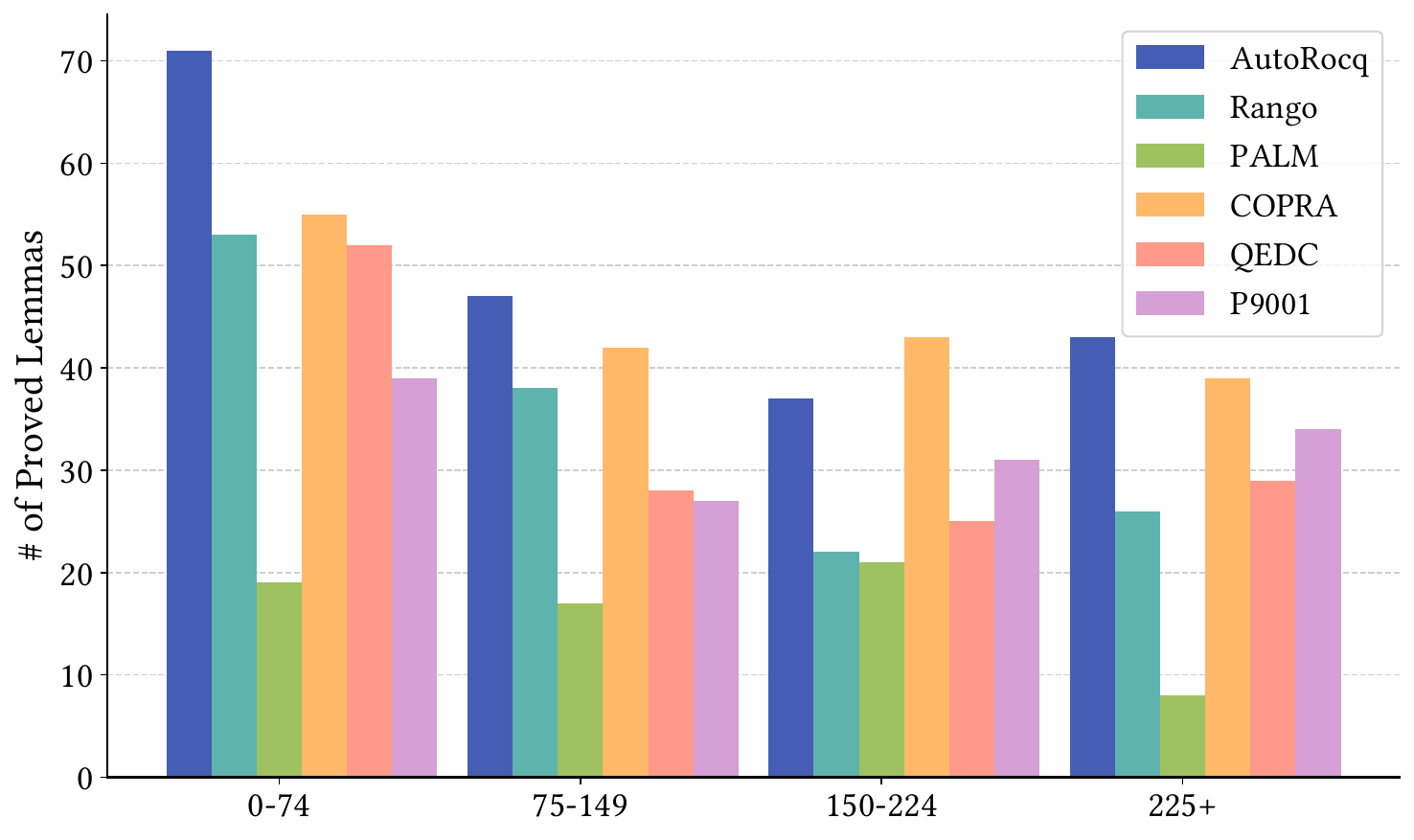}
\label{fig:compl_term}
\vspace{-2em}
\end{minipage}	
}
\hfill
\subfigure[Breakdown by hypothesis count.]{	
\begin{minipage}[t]{0.48\linewidth}
\centering
\includegraphics[width=1\linewidth]{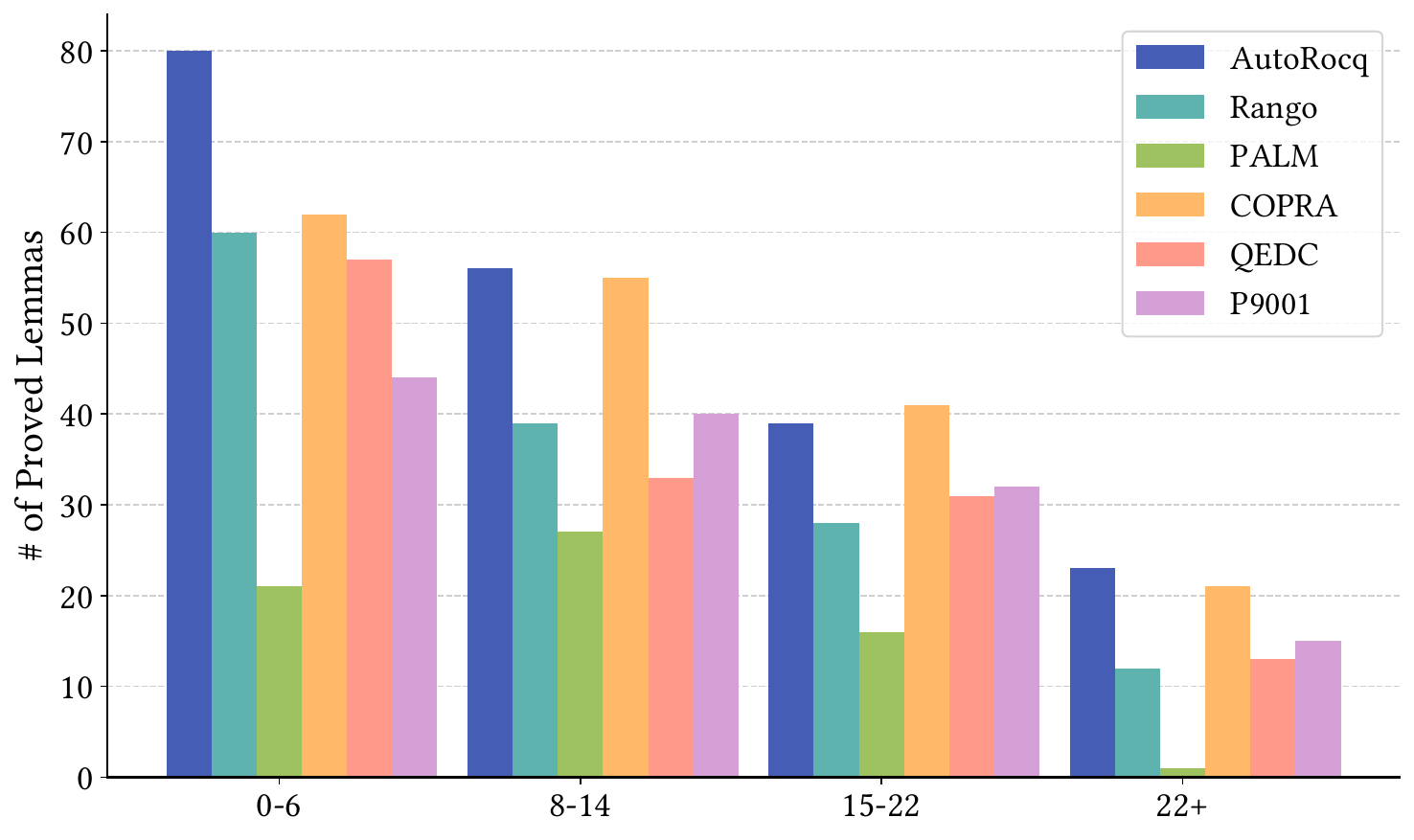}
\label{fig:compl_hypo}
\vspace{-2em}
\end{minipage}	
}
\vspace{-1em}
\caption{\rev{[RQ.2] \# of verification lemmas proved from SV-COMP programs: a breakdown by lemmas' complexity.}}
\label{fig:succ_complexity}
\vspace{-1em}
\end{figure}

To better understand the remarkable efficacy of \ourSol, we further correlate the number of successfully proved SV-COMP lemmas with their complexity.
\autoref{fig:succ_complexity} details the breakdown of successful attempts from different tools, grouped into five complexity buckets.
We present the breakdown for both the term count (Figure~\autoref{fig:compl_term}) and the hypothesis count (Figure~\autoref{fig:compl_hypo}) in the original goal. Buckets on the right correspond to lemmas containing more terms/hypotheses, and thus are more structurally and contextually complicated.
Results reveal an emerging pattern: as the original goal becomes more verbose and context-rich, the number of successfully proved lemmas generally decreases. However, \ourSol retains its relative effectiveness in proving longer, more complex goals, resulting in an almost uniform lead across different complexity levels. 
Among the baselines, \copra scales with increased complexity better, and even manages to prove slightly more lemmas than \ourSol on moderately difficult lemmas (150--224 terms or 15--22 hypotheses). This further demonstrates the advantage of feedback-guided proof generation in program verification.

\begin{figure}[t]
\begin{minipage}{0.48\linewidth}
  \centering
  \includegraphics[width=\linewidth]{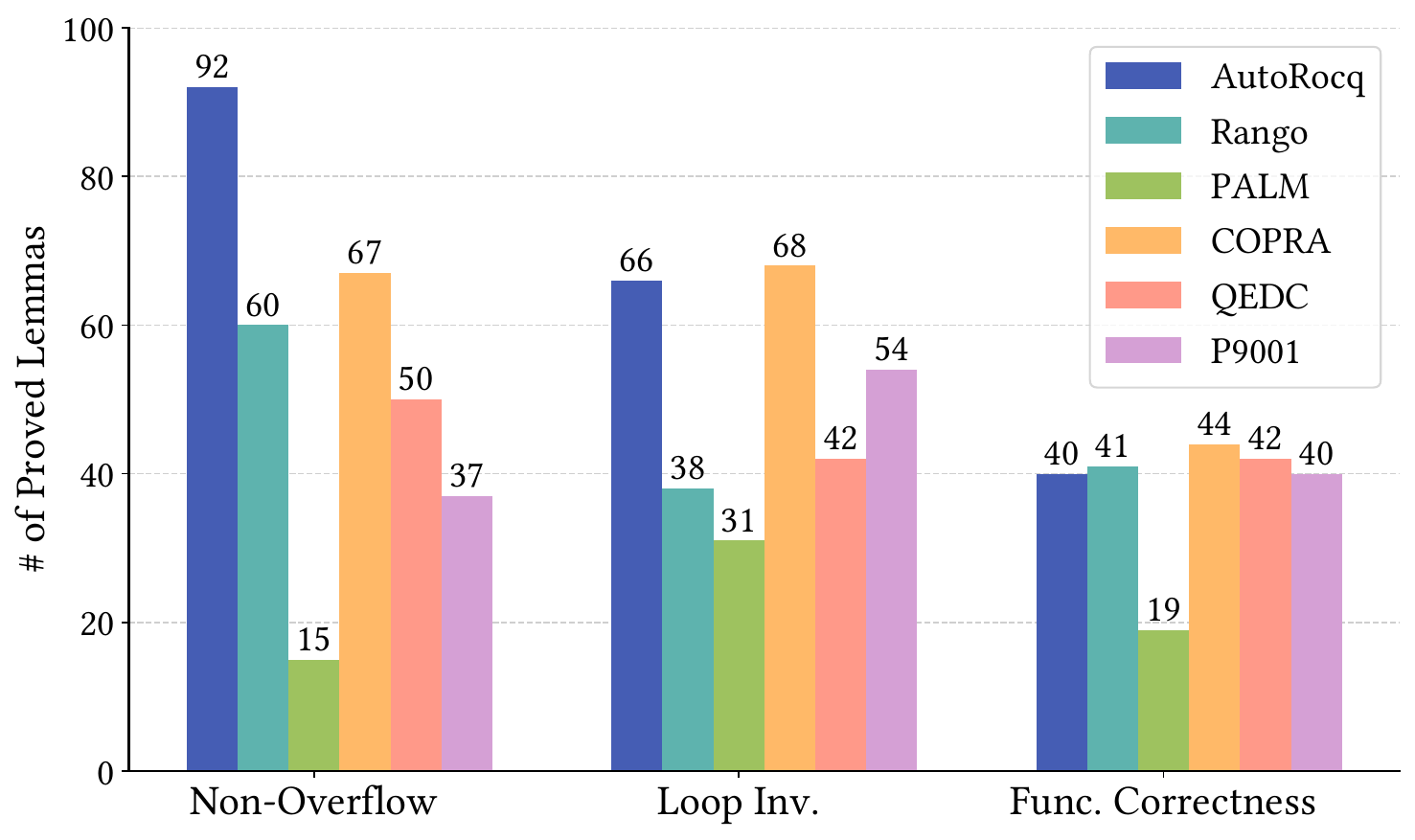}
  \vspace{-2em}
  \captionof{figure}{\rev{[RQ.2] \# of lemmas proved from SV-COMP programs: a breakdown by the category of lemmas.}}
  \label{fig:by_type}
  \vspace{-1em}
\end{minipage}\hfill
\begin{minipage}{0.48\linewidth}
  \centering
  \includegraphics[width=\linewidth]{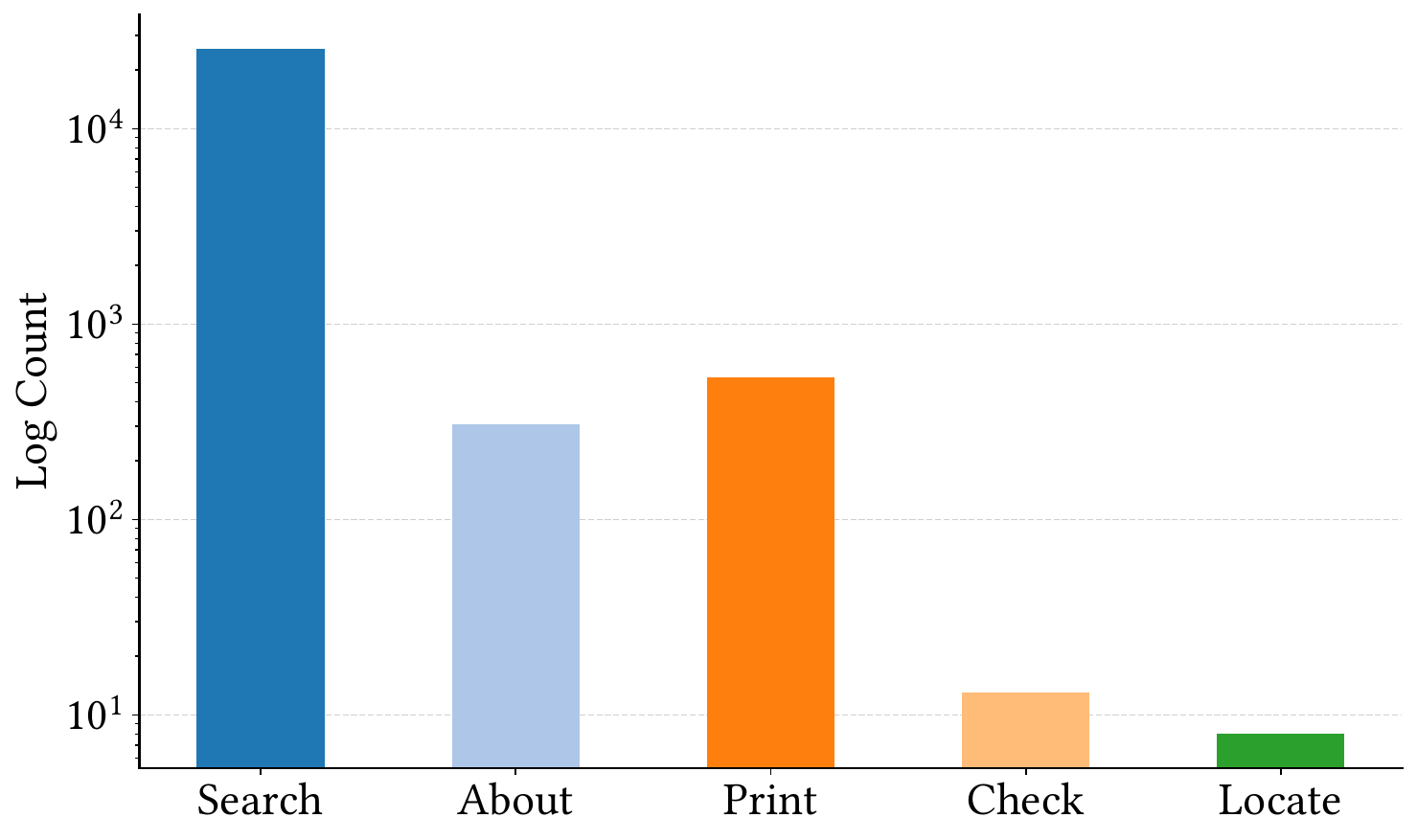}
  \vspace{-2em}
  \captionof{figure}{[RQ.3] The log frequency of invocation for context search commands on SV-COMP programs.}
  \label{fig:search_count}
  \vspace{-1em}
\end{minipage}
\end{figure}

Additionally, we report the breakdown by the categories of lemmas, as shown in \autoref{fig:by_type}.
Results show that \ourSol is particularly effective in proving non-overflow-related lemmas in the program, outperforming the best baselines by 37\%.
On lemmas related to loop invariants, \ourSol is comparable with \copra and outperforms other baselines; whereas on lemmas specifying functional correctness, all approaches except \palm perform comparably.
Results indicate that \ourSol's performance is consistent across different properties, demonstrating its versatility and generality.

We also compare the efficiency of different tools in terms of the time taken to generate a successful proof on the SV-COMP lemmas. 
On average, \ourSol takes 21.3 seconds to generate a successful proof, expending less time than other LLM-based approaches, namely \rango (105.5 seconds), \palm (45.4 seconds), and \copra (49.1 seconds). \ourSol's result is also comparable to \pbot (22.6 seconds). Notably, \qedc finishes extremely fast (5.1 seconds) when it does succeed, thanks to its well-optimized tactic-prediction model for the tactic generation.

We hypothesize that \ourSol's remarkable efficacy and efficiency stem from its agentic access to the proof context, strategic tactic fixing, and effective communication between the LLM and ITPs. These collectively enable the agent to discover and apply relevant facts and lemmas dynamically during the proof search.
These capabilities could be particularly beneficial for complex proof obligations, which often require non-trivial reasoning steps and the application of specific lemmas or theorems.
In the next research question (\autoref{sec:result::rq2}), we evaluate the design decisions of \ourSol to gain a deeper understanding of the system.

\subsection{RQ.3: Design Choices of \ourSol}  \label{sec:result::rq2}

To better understand the interplay between \ourSol's components, we evaluate how different design choices affect the overall effectiveness by implementing 6 variants of \ourSol.
We conduct the evaluation on 70 randomly sampled verification lemmas from SV-COMP programs.
Specifically, we examine the following variant approaches: 
\begin{itemize}[leftmargin=1em,nosep]
    \item \varNoCS: No context search (\autoref{sec:approach::tactic-generation}). Instead, tactics are directly generated through prompting.
    \item \varNoPT: No proof tree awareness (\autoref{sec:approach::proof-tree-interpretation}). Instead, proof states are encoded as plain texts.
    \item \varNoEF, \varNoHM: No error feedback or history manager for the on-the-fly learning (\autoref{sec:approach::feedback-handling}).
    \item \varOne, \varThree, and \varFive: The maximum number of errors encountered before context search. 
    We evaluate three settings of the maximum number of errors encountered before seeking context search. We found that the value 3 is the most effective setting, and is used in \ourSol. This is expected, as more frequent context queries (\varOne) may confuse the agent with unnecessary information, whereas fewer queries (\varFive) may not supply sufficient context.
\end{itemize}

\smallskip
We report the success rates and \ourSol's relative improvement, of different variants in \autoref{tab:rq2_ablation}.
Results show that all variants only prove a fraction of lemmas compared to full-fledged \ourSol.
In particular, \ourSol sees the most significant improvement from the ablative version \varNoCS, which replaces the context search component with a direct request of tactic through prompting.
This indicates that agentic access to the proof context contributes the most to our approach's effectiveness.
Intriguingly, querying for contexts much more frequently (\varOne) or less frequently (\varFive) also affects the results negatively, suggesting that the amount of contexts supplied to the agent is paramount in practice.
Removing other components in our approach, namely the proof-tree representation (\varNoPT) or the feedback mechanisms (\varNoEF~and \varNoHM), each reduces the efficacy moderately by $\sim 10\%-15\%$. We conclude that all components of \ourSol are useful.

To visualize how \ourSol conducts the context search, we summarize the frequency of invocation for different search commands provided by our framework (\autoref{tab:coq-queries}). We plot these frequencies in log scale, as shown in \autoref{fig:search_count}.
Statistics show that {\it Search} is the most frequently used command. Its primacy stems from its unique ability to solve the fundamental problem of discovery: it allows users to find relevant lemmas and theorems without prior knowledge of their names by searching for patterns. In essence, {\it Search} directly facilitates the core intellectual challenge of finding the right fact at the right time.
It also reflects the fact that many proof obligations in our benchmarks require non-trivial reasoning steps, which often hinge on the application of specific lemmas or theorems.
We dig into the specific patterns searched by \ourSol, and find that they often involve complex expressions with multiple operators and operands, e.g., ``{\it (\_ * \_ <= \_ * \_)}~'', ``{\it Z.abs (\_ + \_)}~'', and ``{\it (\_ + \_ <= \_)}~'', which are challenging to write for inexperienced Rocq users.
In short, the above facts highlight that \ourSol's agentic access to the proof context through invocation of {\it Search} and other queries is pivotal in its efficacy.

\begin{resultbox}
{\bf Answer to RQ.3}: Each component of \ourSol contributes to its overall effectiveness. \ourSol's agentic context search enhances its efficacy most significantly.
\end{resultbox}

\begin{table}[t]
 \centering
 \caption{[RQ.3] Success rates of different variants of \ourSol on proving sampled verification lemmas. %All components contribute to the overall effectiveness, with context search (\varNoCS) being the most significant.
 }
 \label{tab:rq2_ablation}
 \vspace{-0.5em}
 \small
 \begin{tabular}{l | c | c c c c | c c }
 \toprule
     & \ourSol & \varNoCS & \varNoPT & \varNoEF & \varNoHM & \varOne & \varFive \\
 \midrule
    Success Rate	& 44.3\% &	30.0\% &	40.0\% & 38.6\%	& 41.4\% & 37.1\% & 38.6\% \\
 \hline
    Relative Improv. & -- & +47.6\% &	+10.7\% & +14.8\% & +6.9\% &	+19.2\%	& +14.8\% \\
 \bottomrule
 \end{tabular}
 %\vspace{-1em}
\end{table}

\subsection{RQ.4: Comparison to Human-written Proofs} \label{sec:result::rq3}

Since there is no ground truth for the lemmas extracted from SV-COMP programs, we cannot directly compare the proofs synthesized by \ourSol with human-written proofs.
In this subsection, we closely examine a proof synthesized by \ourSol, and compare it to a manually crafted proof by a Rocq expert. 
The expert has more than six years of formal verification experience and one year of practical experience with Rocq. 
We use the same theorem {\tt wp\_goal} as presented in \autoref{fig:motivating-example}(a), and report the human-written proof in \autoref{fig:proof-cmp}.

To construct the proof in Figure \autoref{fig:proof-human}, the expert had first to recognize that the problem's core reduces to the mathematical principle that, for any integer $a$, $a^2 \geq 100$ implies $|a| \geq 10$. This insight is non-trivial, as it requires moving beyond a direct case-based enumeration of values to grasp the underlying structural inequality. Only after achieving this conceptual leap could the expert elegantly bifurcate the problem based on the sign of {\tt i1} and construct the two custom, symmetric helper lemmas (lines 9--15 in Figure \autoref{fig:theorem}) to complete the solution.
Crafting this proof takes the Rocq expert 20 minutes of time.

In contrast, let us consider again the proof generated automatically by \ourSol in lines 9--23 in \autoref{fig:motivating-example}(a). 
It took \ourSol only 156.7 seconds to synthesize a proof for the same lemma.
At a high level, it achieves the same realization as the human prover, and focuses its proof on an explicit, enumerative case analysis for the value of {\tt i1}.
It also quickly connects the goal to properties of absolute values ({\tt Z.abs}).
Crucially, \ourSol leverages two key lemmas from the global context to assist its reasoning on absolute values, namely {\tt Z.abs\_le} at line 12 and {\tt Z.square\_le\_mono\_nonneg} at line 19.
This not only allows it to prove the goal successfully, but also proves it in noticeably fewer steps than the human expert.
Our log reveals that \ourSol autonomously invokes multiple context searches during its proving. Specifically, searching for the pattern ``{\it (\_ * \_ <= \_ * \_)}~'' surfaces the crucial lemma used at line 19, which eventually helps \ourSol to succeed. Unsurprisingly, \emph{no} other baseline tools are able to prove this lemma due to their lack of agency.

\begin{figure}[t]
%\vspace{-0.5em}
\centering
\subfigure[{\tt wp\_goal} in \autoref{fig:motivating-example}(a) with human helper lemmas.]{
\begin{minipage}[c]{0.48\linewidth}
\centering
\tiny
\begin{lstlisting}[language=Coq, basicstyle=\tiny\ttfamily]
Theorem wp_goal :
  forall (i:Numbers.BinNums.Z) (i1:Numbers.BinNums.Z),
  let x := (i1 * i1)%Z in ((-9%Z)%Z <= i1)%Z -> 
  ((-9%Z)%Z <= i)%Z -> (i <= 9%Z)%Z -> (i1 <= 10%Z)%Z -> 
  (i <= 10%Z)%Z ->((-2147483648%Z)%Z <= x)%Z -> 
  (100%Z <= x)%Z -> (x <= 2147483647%Z)%Z -> is_sint32 i1 -> 
  is_sint32 i -> (i1 = 10%Z).

(* Helper lemmas written by the human prover*)
Lemma square_le_mono_nonneg :
  forall a b, 0 <= a -> 0 <= b -> a * a <= b * b -> a <= b.
Proof. intros a b Ha Hb Hsq. nia. Qed.
Lemma square_le_mono_neg :
  forall a b, a < 0 -> b < 0 -> a * a <= b * b -> b <= a.
Proof. intros a b Ha Hb Hsq. nia. Qed.
\end{lstlisting} % Replace with the path to your second code file
\end{minipage}	\label{fig:theorem}
}
\hfill
\subfigure[Main proof script written by a human Rocq expert.]{	
\begin{minipage}[c]{0.48\linewidth}
\centering
\tiny
\begin{lstlisting}[language=Coq, basicstyle=\tiny\ttfamily]
Proof.
  intros. subst x. assert (Hcases: 0 <= i1 \/ i1 < 0) by lia.
  destruct Hcases as [Hge0 | Hlt0].
  - assert (H10: 0 <= 10) by lia.
    pose square_le_mono_nonneg. 
    specialize (l 10 i1 H10 Hge0). 
    simpl in l.
    specialize (l H5). lia.
  - pose square_le_mono_neg.
    specialize (l (-10) i1).
    assert (Hn10: -10 < 0) by lia.
    specialize (l Hn10 Hlt0). 
    simpl in l.
    specialize (l H5). lia. 
Qed.  

\end{lstlisting} % Replace with the path to your second code file
\end{minipage}	\label{fig:proof-human}
}
\vspace{-1em}
\caption{[RQ.4] Proof for {\tt wp\_goal} in \autoref{fig:motivating-example}(a), written by a human expert prover in 20 minutes.}
\vspace{-1em}
\label{fig:proof-cmp}
\end{figure}

\begin{resultbox}
{\bf Answer to RQ.4}: With the help of context queries, \ourSol is able to prove challenging lemmas, sometimes in even fewer steps than human experts. It also takes much less time.
\end{resultbox}

\subsection{Threats to Validity} \label{sec:threat}

% In this section, we discuss potential threats and limitations in our approach and evaluation.

\subsubsection*{\bf Trustworthiness and Cost of LLMs.} LLMs may give incomplete or factually wrong responses to questions.
As a result, verifying programs with untrusted LLMs may seem off-putting at first glance. However, we note that the certificate essentially comes from the trusted kernel of Rocq, which ensures that only derivations conforming to the formal rules of the proof assistant are accepted \cite{coq-kernel}. 
Therefore, all the generated proof scripts are \emph{true positives} and are trustworthy, as any erroneous or misleading outputs from the LLMs will simply fail to be verified.
Throughout our experiments, it takes \$1.22 on average to prove a single lemma without token caching. We deem this cost acceptable, compared to the costly human involvement in lemma-proving activities.

\smallskip
%\subsubsection*{\bf Data Leakage.}
{\bf Data Leakage.}
LLMs are trained on a large corpus of data; thus, they may have been exposed to open-sourced Rocq projects. As such, LLM-based approaches, including \ourSol, may report inflated results on \coqgym, which comprises only lemmas and proofs in the public domain. 
In our evaluation, we mitigate this risk by including obligations from SV-COMP programs.
These lemmas are extracted by us automatically (see more details in \autoref{sec:lemma_extraction}), and their ground-truth proofs are not available. In fact, it is unclear if these lemmas are provable at all until we find a correct proof, so we believe the risk of data leakage is minimal.

\smallskip
%\subsubsection*{\bf Lemma Selection.}
{\bf Lemma Selection.} 
In our evaluation, we include established benchmarks and verification targets, namely \coqgym and SV-COMP programs. The lemmas are selected systematically as detailed in \autoref{sec:benchmark_selection}.
We study (un)reachability and overflow freedom as examples of correctness and safety properties, respectively, due to their universality. 
We also study the loop invariants proposed by LLMs.
However, certain types of programs or properties may pose systematic challenges to our approach that we are unaware of. We plan to investigate more types of programs (e.g., multi-threaded) or properties (e.g., memory safety) in future work.

\smallskip
\textbf{Impact of LLMs Used.}
Since \ourSol adopts a general-purpose LLM (i.e., GPT-4.1) while some baseline approaches adopt custom ones, we want to study the impact of the underlying model. We conducted an additional comparison between \ourSol and a variant of \rango (the best-performing baseline with a custom model) named \rango-GPT-4.1.
In this variant, the fine-tuned LLM was replaced with GPT-4.1. Experimental results show that \rango-GPT-4.1 can only prove 11.43\% of the verification lemmas used in Section~\ref{sec:result::rq2}, whereas \ourSol proves 44.29\%. These results strongly indicate that \ourSol's relative efficacy over competing tools comes from the agentic layer rather than from a powerful choice of LLM. 

\smallskip
\textbf{Impact of Rocq Versions.}
During evaluation, we used a different Rocq 8.12 for \palm, which is different from other comparative tools (i.e., Rocq 8.18), raising concerns about how different Rocq versions may affect the experimental results. 
In theory, Rocq version differences do not materially affect the comparison between the \palm and \ourSol, as most of the version changes concern syntax, parsing, or library organization, rather than the proof engine that our method relies on. Practically, the impact would be minimal, as \ourSol dynamically retrieves and queries lemmas from the current Rocq environment and thus does not depend on version-specific assumptions.

\section{Case Study: Verifying Linux Kernel Modules} \label{sec:linux}

\begin{table}[!t]
 \centering
 \caption{[Cast Study] Verifying functional correctness in Linux kernel modules: \# of lemmas proved, and the average time (in seconds) and steps (in \# of tactics) expended on proved lemmas. %\ourSol proves 12 (20\%) of all 60 lemmas. With \coqhammer integration, \ourSol-w/H proves 18 (30\%) of them. 
 }
 \vspace{-1em}
 \label{tab:linux}
 \small
 % \resizebox{\linewidth}{!}{
 \begin{tabular}{l | c c c c c c | c }
 \toprule
    Tools & \ourSol & \rango & \palm & \copra & \qedc & \pbot & \rev{\ourSol-w/H} \\
 \midrule
    Proved? (max. 60) & \bf{12} & 3 & 10 & \rev{7} & 3 & 2 & \bf{\rev{18}} \\
    Avg. Time  & 29.9 & 96.2 &	39.4 &\rev{65.1} & \bf{2.0} & 160.8 & \rev{65.9} \\
    Avg. Steps & \bf{14.3} & 32.0 & -- & \rev{50.0} & 32.0 & 979.0 & \rev{16.7} \\
 \bottomrule
 \end{tabular}
 % }
 \vspace{-1em}
\end{table}

\ourSol has demonstrated remarkable efficacy in automatically synthesizing rigorous proofs for lemmas related to smaller programs or a mathematical context.
In this case study, we assess if \ourSol can handle the complexity of real-world software.
To this end, we build on earlier efforts \cite{volkov2018lemma, efremov2018deductive} to reason about properties in the Linux kernel. 
Specifically, we examine 60 lemmas that formalize functional correctness in Linux kernel code. These lemmas reason about the correctness of intricate program behaviors, and come from various source files, such as memory management (e.g., \texttt{memmove}) and utilities such as string operations (e.g., \texttt{strcpy}) and type conversion (e.g., \texttt{hex2bin}).
These lemmas tend to be significantly more involved and may contain up to hundreds of terms.

We report the statistics in \autoref{tab:linux}. 
Results show that \ourSol is the most effective, being able to synthesize the proofs for 12/60 (20\%) lemmas.
\copra manages to prove 7 lemmas as well.
In contrast, \rango, \qedc, and \pbot perform poorly on these tasks, which only manage to succeed in 3, 3, 2 lemmas, respectively.
We also note that \qedc runs exceptionally fast when it is able to generate a proof, being 15x faster than \ourSol, which is the second fastest. This suggests that \qedc's search heuristic performs extremely well on certain cases. Nonetheless, such cases seem rare when verifying real, complex software, as it only succeeds on 3 lemmas.

Notably, \palm also achieves remarkable effectiveness and proves 10/60 lemmas. Upon manual examination, we note that \emph{all} successful proofs generated by \palm rely on heavy use of \coqhammer~\cite{coqhammer}, a plug-in that directly discharges remaining subgoals to automated theorem provers (ATPs). \coqhammer is invoked by \palm as a last resort when \emph{none} of its deterministic tactics repairs or resolves the encountered errors. 
Based on this finding, we create a new variant, \ourSol-w/H, by augmenting our approach with basic \coqhammer integration.
Specifically, the hammer is invoked with its default settings \emph{once} per proof state upon the first failed tactic application. 
Such integration yields significant efficacy gains, with 18 lemmas (30\%) proved, indicating that access to additional tools could further enhance the capabilities of our proof agent.

\begin{resultbox}
    \ourSol can be applied to verify properties in complex software such as the Linux kernel. Access to external automated theorem provers (e.g., \coqhammer) further improves \ourSol.
\end{resultbox}

\section{Related Work}  \label{sec:related-work}

\subsection{Deductive Program Verification}

Deductive verification formally verifies the correctness of software systems by extracting and proving a collection of mathematical proof obligations from both the program and its specification \cite{filliatre2007krakatoa}.
The automation of the verification process is largely underpinned by verification-oriented languages, including \framac~\cite{DBLP:journals/fac/KirchnerKPSY15}, \dafny~\cite{DBLP:conf/lpar/Leino10}, \verus~\cite{DBLP:journals/corr/abs-2303-05491}, and \viper~\cite{DBLP:conf/vmcai/0001SS16}, which provide an end-to-end verification flow that integrates specification and proof directly into the programming workflow through the use of annotations. 
These tool chains generate proof obligations automatically, thereby reducing the problem of program verification to theorem proving. 
The proof obligations are sent to either Automated Theorem Provers (ATPs) or Interactive Theorem Provers (ITPs) to verify.
ATPs such as \coqhammer~\cite{coqhammer} attempt to discharge the obligations automatically through constraint solving~\cite{de2008z3, barbosa2022cvc5}, while ITPs such as Isabelle/HOL~\cite{paulson1994isabelle}, Lean~\cite{DBLP:conf/cade/Moura021}, and Rocq~\cite{coq1996coq} allow users to interactively construct proofs using tactics.
Due to ATPs' limited capabilities in handling complex proofs \cite{nieuwenhuis2007challenges} and occasional correctness issues~\cite{smt-bugs-fse20,smt-yingyang-pldi20}, ITPs have become the \textit{de facto} standard for verifying sophisticated software systems.
Indeed, many pioneering efforts have successfully verified critical software systems~\cite{DBLP:conf/sosp/KleinEHACDEEKNSTW09, DBLP:conf/osdi/ChajedTTKZ22, leroy2016compcert, verified_db}, network protocols~\cite{DBLP:conf/sosp/ZhengLXSLHWMLCZ23}, and microprocessor designs~\cite{DBLP:journals/dt/JonesOSAM01} with ITPs. 
However, ITPs require significant human effort to guide the proof construction, which limits their scalability and applicability.
\ourSol specifically focuses on reducing the manual effort in proving complex residual obligations, thus advancing towards an automatic, end-to-end verification workflow. 
As such, it is related but complementary to prior work automating other aspects of the workflow, including inference of specifications~\cite{endres2024can} and loop invariants~\cite{DBLP:conf/cav/SiNDNS20, si2018learning, DBLP:journals/tmlr/LoughridgeSACS025, learning-invariants}, discovery of helper lemmas~\cite{sivaraman2022data}, and direct synthesis of verified methods \cite{DBLP:journals/pacmse/MisuLM024}.

\subsection{Automated Techniques in Theorem Proving}
Our work is closely related to the substantial research devoted to automating mathematical theorem proving~\cite{lisurvey,alpha-proof}, spanning the complete pipeline from premise selection and tactic generation to proof search and recovery mechanisms. 
Our focus on proving program-derived lemmas leads to unique design choices in \ourSol, which we detail below.

%\smallskip
{\bf Premise Selection.} Premise selection is the task of identifying relevant lemmas, definitions, or axioms from a large library to assist in proving a given theorem \cite{yang2023leandojo, mikula2023magnushammer}.
Early approaches treat premise selection as a binary classification task, where the goal is to predict whether a given premise is relevant to the theorem at hand \cite{irving2016deepmath}.
More recent approaches such as \coqhammer~\cite{coqhammer}, \palm~\cite{lu2025palm}, and \rango~\cite{rango-icse25} leverage machine learned weights or lexical similarity measures~\cite{sparck1972statistical} to rank premises statically.
In contrast, \ourSol adopts on-demand premise retrieval with the support of pattern-based context search.

%\smallskip
{\bf Tactic Generation.} Tactic generation is the core component in interactive theorem proving, where a proof step is proposed to transform the current proof state \cite{lisurvey}.
Early works rely on hand-crafted heuristics and domain-specific knowledge to guide generation \cite{leroy2016compcert,certiOS,DBLP:conf/sosp/KleinEHACDEEKNSTW09}. 
More recently, machine learning techniques have been applied to learn step-wise generation strategies from large corpora of human-written proofs \cite{DBLP:conf/nips/WuNWD21, alpha-proof, blaauwbroek2020tactician, yang2019coqgym, sanchez2020generating,first2020tactok,rango-icse25}.
\ourSol follows the same paradigm and queries the underlying LLM at each step with a tree-based proof state representation. 

%\smallskip
{\bf Proof Search.}
Proof search concerns the \emph{sequence} of tactic applications, and has been a long-standing challenge due to an inherently large search space and sparse rewards \cite{proof-search,lample2022hypertree}. 
Whole-proof generation approaches such as \palm~\cite{lu2025palm} circumvent the search process by predicting the entire tactic sequence in a single pass. 
Step-by-step proving techniques conventionally employ beam search to sample multiple tactic predictions with breadth-first \cite{bansal2019holist}, depth-first \cite{yang2019coqgym}, or best-first heuristics \cite{polu2020generative, rango-icse25} to traverse the search space. 
Recent works have leveraged other exploration mechanisms such as reinforcement learning~\cite{DBLP:conf/nips/WuNWD21, sanchez2024qedcartographer} and in-context learning~\cite{thakur2024context}.
\ourSol adapts the search strategy through its agency, enabled by timely feedback, high-level interpretation of the proving progress, and on-the-fly learning from successful histories.

%\smallskip
{\bf Recovery Mechanisms.} 
Due to the trial-and-error nature of theorem proving, proof automation systems need to effectively recover proof attempts and explore alternative strategies \cite{lu2025palm,thakur2024context}.
Yet this issue is often not addressed explicitly. Rather, a large body of work takes only binary signals from ITPs and simply retries tactic generation in case of failures~\cite{rango-icse25, sanchez2024qedcartographer}. Others, such as \copra~\cite{thakur2024context} and \proverbot~\cite{sanchez2020generating}, are able to restore to an earlier proof state if necessary.
\palm~\cite{lu2025palm}, on the other hand, performs opportunistic tactic repairs and delegates the open subgoal to an ATP~\cite{coqhammer} as the last resort.
In contrast, \ourSol exploits detailed diagnostic errors from ITPs to rectify failed tactic applications, and handles persistent errors through autonomous context enrichment.

\subsection{LLM for Software Quality Assurance}

Besides formally verifying correctness through theorem proving as is done in \ourSol, a wide spectrum of techniques for software quality assurance have benefited from recent advances in Large Language Models \cite{hou2024large,fan2023large}. 
For example, model checking techniques also offer strong correctness guarantees by encoding a software system as a finite state machine and iteratively checking its states and transitions. LLMs have been employed to automate the construction of both the model \cite{zuo2025pat} and the formal specification \cite{llm-model-checking}. 
Testing techniques, on the other hand, try to prove the \emph{presence} of bugs by synthesizing test cases that trigger unintended behaviors, and have benefited significantly from LLMs' capability in input generation \cite{meng2024large, deng2024large, xia2024fuzz4all} and program understanding \cite{cottontail-sp26, ConcoLLMic}.
Static analysis leverages LLMs to identify code issues and vulnerabilities faster and earlier in the development process \cite{li2024enhancing, wu2024advscanner}. 
Recent work also applies agentic hybrid (static–dynamic) analysis to security tasks such as cross-platform rule conversion \cite{xu2026}.
While each technique above presents unique trade-offs, deductive verification stands out for its mathematical rigor and stringent guarantees on correctness, making it particularly well-suited for safety-critical systems~\cite{efremov2018deductive}. 

\section{Perspectives} 

We present an agentic proving system that automatically generates machine-checkable certificates for proof obligations.
At its core, our agent is an autonomous process that retrieves relevant contexts on demand, incorporates feedback from the proof assistant, and generates tactics adaptively -- all guided by interpreting the proof derivation tree. 
Together with a lemma extraction process, our agent can achieve effective push-button verification that takes source code in C (i.e., SV-COMP programs and Linux kernel modules) and automatically generates proofs without any human effort.

As AI-generated code becomes increasingly prevalent in software development, the need for automated verification becomes more pressing. Our work demonstrates that LLM agents can bridge the gap between automatic code generation and formal verification, moving closer to the vision of trusted automatic programming. The agentic nature of our approach suggests a paradigm shift where verification tools act as intelligent collaborators rather than passive validators, capable of autonomously navigating complex proof spaces and adapting to diverse verification challenges. This represents a crucial step toward making formal verification accessible for real-world software systems.
We can thus move closer to the vision of AI-based Verification and Validation (V \& V) of AI-generated code. This will alleviate the problem of humans reviewing AI-generated code.

\begin{acks}
We would like to sincerely thank all the anonymous reviewers for their valuable comments.
This work was partially supported by a research project ``AI for Program Reasoning'', project AI4SCH-2025-0084 under AI for Science program of National Research Foundation (NRF) Singapore. 
Any opinions, findings and conclusions, or recommendations expressed in this material are those of the author(s) and do not reflect the views of National Research Foundation, Singapore.
\end{acks}

\bibliographystyle{ACM-Reference-Format}

\section*{Data Availability}

We release \ourSol, including its implementation, benchmarks, and replication instructions, for academic use at the following link:
\url{https://github.com/NUS-Program-Verification/AutoRocq}.

\bibliography{sections/ref}

\end{document}
\endinput
%%
%% End of file `sample-acmsmall-conf.tex'.